\documentclass[
 reprint, % Enables 2-column mode
 amsmath, % Enables common math environments, e.g. align and gather
 aps, % Sets up proper APS styles e.g. for bibliography
 pra, % Journal-specific formatting settings 
 showkeys, % For displaying keywords
 superscriptaddress % For affiliation numbering
]{revtex4-2}

\usepackage{hyperref} % Enables clickable links, e.g. to citations or figures
\usepackage{graphicx} % Enables inclusion of figure files
\usepackage{orcidlink} % Enables green orcid links next to author names
\usepackage[caption=false]{subfig} % Enables subfigures
\usepackage{array} % Enables text wrapping and centering in tables
\usepackage{enumitem} % Enables spacing control around items of lists
\usepackage[ruled, linesnumbered]{algorithm2e} % Enables pseudocode environment
\usepackage{braket} % For bra-ket notation commands
\usepackage{setspace} % For spacing control

\hypersetup{
    colorlinks,
    linkcolor=blue,
    citecolor=blue,
    urlcolor=blue
} % Makes links blue and removes rectangles around them

\newcolumntype{C}[1]{>{\centering\arraybackslash}p{#1}} % Defines fixed-width horizontally-aligned column type. Width is given by the argument. E.g. \begin{tabular}{|C{3cm}|C{2.5cm}|}

\setlist{itemsep=-5pt, topsep=5pt} % Reduces between-item list spacing and adds spacing between list and surrounding text

\begin{document}
% \doublespacing

\title{Benchmarking Hybrid Quantum-Classical Algorithms for Power Grid Optimization Problems}

\author{Igor Gaidai\orcidlink{0000-0002-3950-3356}}
\email{igor-gaidai@utc.edu}
\affiliation{Department of Physics and Astronomy, University of Tennessee, Chattanooga, USA}
\affiliation{Quantum Center, University of Tennessee, Chattanooga, USA}

\author{Rick Mukherjee}
\email{rick-mukherjee@utc.edu}
\affiliation{Department of Physics and Astronomy, University of Tennessee, Chattanooga, USA}
\affiliation{Quantum Center, University of Tennessee, Chattanooga, USA}

\begin{abstract}
Alternating Current Optimal Power Flow Unit Commitment (AC-OPF-UC) is a difficult mixed-integer nonlinear optimization problem that combines binary generator commitment decisions with nonconvex continuous AC power-flow constraints.
In this work, we investigate whether hybrid quantum-classical variational algorithms can improve the solution of single-period AC-OPF-UC relative to classical approaches.
To the best of our knowledge, this is the first study to directly evaluate quantum or hybrid quantum-classical algorithms for single-period AC-OPF-UC with explicit binary commitment variables and the full nonlinear AC power-flow formulation.
We consider two candidate algorithms for improving AC-OPF-UC solution quality relative to purely classical methods on ideal quantum hardware.
The first applies QAOA directly to a fully discretized formulation of the problem, with equality and inequality constraints incorporated through penalty terms and slack variables.
Although conceptually straightforward, this approach requires a prohibitively large number of qubits even for small instances.
The second, qubit-efficient approach encodes only the binary generator status variables on a quantum computer, while optimizing the continuous power-flow variables classically for each sampled bitstring.
We benchmark this method on randomly generated AC-OPF-UC instances with 5 to 13 generators and compare it against SCIP, SMAC, and uniform random sampling.
Our simulations show that the qubit-efficient hybrid method does not outperform uniform sampling.
These results suggest that in order to establish potential advantage of the variational hybrid strategy considered here over the best classical algorithms, if any, much larger system sizes (24+ generators) need to be tested, which is beyond our computational capacity.
Alternatively, different approaches, such as quantum versions of branch-and-bound methods, may be more promising.
\end{abstract}


\maketitle

\section{Introduction}
\label{sec:intro}

Power grid optimization is a class of optimization problems concerned with cost-efficient energy production, distribution of power and reliability in power networks.
One such problem that we focus on in this work is called Alternating Current Optimal Power Flow Unit Commitment (AC-OPF-UC) problem.
Roughly speaking, the problem asks to find node voltages, phase angles and power output levels of generators in the network, such that the consumer load demand is satisfied, power production cost is minimized and voltages are kept as close to nominal as possible. 
The exact problem statement can be found in Section~\ref{sec:theory}.
This problem is notoriously difficult, combining hard combinatorial optimization over binary variables with a non-convex optimization of continuous variables.
Finding any feasible solution, much less an optimal one, for even a simpler variation of the problem (AC-OPF, without the binary variables) is known to be strongly NP-hard. \cite{bienstock2019strong}

Classical methods for the AC-OPF-UC problem are best understood as different ways of managing two coupled sources of difficulty: binary generator commitment decisions and nonlinear AC power-flow constraints.
The most direct approach is to preserve both parts of the model and solve the problem as a nonconvex Mixed-Integer Nonlinear Programming (MINLP) or Mixed-Integer Quadratically Constrained Quadratic Programming (MIQCQP) problem.
This class of methods can provide the strongest solution guarantees, typically in the form of feasible incumbent solutions together with primal-dual optimality gaps.
However, these guarantees come at high computational cost, since the mixed-integer search can grow exponentially in the worst case.
As a result, direct global methods are most useful for small and moderate benchmark systems, but do not by themselves resolve the scalability problem \cite{castillo2016unit, liu2018global}.

A common way to improve tractability is to simplify the network model before optimization.
For example, replacing the AC network with a linearized Direct Current (DC) approximation leads to Mixed-Integer Linear Programming (MILP) or Mixed-Integer Quadratic Programming (MIQP) formulations that are much easier to solve.
The resulting schedules can then be checked or repaired using AC feasibility tests.
This approach is computationally attractive, but it generally cannot guarantee feasibility or optimality for the original AC problem because voltage magnitudes, reactive power, losses, and nonlinear branch-flow constraints are not fully represented \cite{baker2021solutions, tuncer2022misocp}.

Between these two extremes, much of the recent literature uses convex relaxations that retain more of the AC network structure than DC approximations while remaining more tractable than the original nonconvex formulation.
Semidefinite Programming (SDP), Second-Order Cone Programming (SOCP), and Mixed-Integer Second-Order Cone Programming (MISOCP) formulations replace the original nonconvex constraints with tractable lower-bounding problems \cite{low2014convex, molzahn2019survey}.
When the relaxation is exact, the solution of the relaxed problem is also globally optimal for the original nonconvex problem, so the relaxation provides a certificate of global optimality \cite{low2014convex, molzahn2019survey}.
When the relaxation is not exact, it can still provide useful lower bounds, but the corresponding relaxed solution may not be feasible for the original AC formulation and may require feasibility restoration or local AC-OPF polishing \cite{molzahn2019survey}.
These relaxations are often combined with decomposition, which separates the problem into smaller continuous and mixed-integer subproblems and improves scalability in AC network-constrained UC applications \cite{constante2021ac, tuncer2022misocp, zohrizadeh2018sequential, parker2024managing}.
This direction has produced strong practical results on standard benchmark systems, but exactness and feasibility recovery remain instance-dependent.

Despite these advances, the problem remains difficult for large system sizes, which makes it a natural candidate for investigation of whether quantum computing can alleviate its difficulty. 
Quantum and hybrid quantum-classical approaches to power-grid optimization have so far been applied mainly to simplified subproblems, as opposed to the bigger AC-OPF-UC with binary commitment variables and the full nonlinear AC power-flow constraints considered here (for a single time period).
For power-flow and OPF models, one line of work uses quantum linear-system solvers, such as HHL \cite{harrow2009quantum} or variational quantum linear solvers \cite{bravo2023variational}, either directly for power-flow equations \cite{eskandarpour2020quantum, saevarsson2022quantum} or inside Newton and interior-point iterations \cite{neufeld2024hybrid, amani2023quantum, amani2025optimal, hafshejani2025quantum}.
In these approaches, the quantum computer accelerates the repeated solution of linear systems arising from the KKT or Newton equations.
This gives a plausible asymptotic motivation, but it is not directly a solver for AC-OPF-UC, since by itself it does not handle binary commitment decisions, branch-and-bound search, or the nonconvex mixed-integer structure of the full problem.
Existing demonstrations are also mostly small, simulated, or limited to DC-OPF, AC-OPF, or power-flow equations, and they generally reproduce classical interior-point behavior without showing a practical speedup \cite{eskandarpour2020quantum, saevarsson2022quantum, neufeld2024hybrid, amani2023quantum, amani2025optimal, hafshejani2025quantum}.

A second direction encodes unit commitment as a QUBO or Ising Hamiltonian and applies QAOA \cite{farhi2014quantum}, quantum annealing \cite{rajak2023quantum}, or related hybrid methods \cite{ajagekar2019quantum, koretsky2021adapting, salgado2024hybrid, hasanzadeh2026survey}.
Here the quantum computer is used to search over binary generator-status variables.
This is more naturally connected to the commitment part of AC-OPF-UC, but it does not straightforwardly include continuous AC voltages, reactive powers, nonlinear power-flow equations, or line-current constraints.
Those variables must either be discretized, which greatly increases qubit requirements, or optimized classically in a decomposition scheme.
Related hybrid decompositions have also been studied for DC-OPF-UC and distributed UC, where quantum routines are embedded inside ADMM, Lagrangian relaxation, or surrogate Lagrangian frameworks \cite{mahroo2022hybrid, magar2024dc, nikmehr2022quantum, feng2022novel}.
These methods are closer in spirit to AC-OPF-UC, but most rely on DC power flow, simplified network constraints, or separable subproblems, so their extension to the full nonconvex AC formulation is not automatic.

Overall, the studies reviewed above suggest that quantum methods may be useful as components inside hybrid decompositions, especially for linear algebra \cite{amani2025optimal, hafshejani2025quantum} or binary search subroutines \cite{ajagekar2019quantum, koretsky2021adapting, salgado2024hybrid, hasanzadeh2026survey, mahroo2022hybrid, magar2024dc, nikmehr2022quantum, feng2022novel}.
However, current evidence does not establish a practical advantage over strong classical baselines for realistic grid-optimization problems, and no existing approach directly resolves the combined difficulty of binary commitment and nonlinear AC feasibility.
To the best of our knowledge, no other work studied whether quantum algorithms are viable for AC-OPF-UC, how much improvement can they bring and at what cost.
Thus, the goal of this paper is to close this gap and answer these questions.

\section{Background}
\label{sec:background}

\subsection{Time-Dependent Picture}

Consider an arbitrary circuit element in an Alternating Current (AC) circuit.
Voltage at the element oscillates sinusoidally:
\begin{equation}
    v(t) = \sqrt{2} \bar{V} \cos(\omega t + \phi_v)
\end{equation}
where $\bar{V}$ is the root mean square (RMS) voltage, $\omega$ is oscillation frequency, $t$ is time, and $\phi_v$ is a phase.

Typically, current through the element also oscillates sinusoidally with the same frequency, but may, in general, have a different phase (because of reactive circuit elements, such as inductors and capacitors):
\begin{equation}
\label{eq:current}
    i(t) = \sqrt{2} \bar{I} \cos(\omega t + \phi_i)
\end{equation}
where $\bar{I}$ is RMS current.

Power of a circuit element is a measure of the rate of work that element performs.
Power can be calculated as a product of voltage and current:
\begin{equation}
\label{eq:power}
\begin{aligned}
    p(t) &= v(t) i(t) = 2 \bar{V} \bar{I} \cos(\omega t + \phi_v) \cos(\omega t + \phi_i) \\
    &= \bar{V} \bar{I} (\cos(\phi_v - \phi_i) + \cos(2 \omega t + \phi_v + \phi_i)) \\
    &= \bar{V} \bar{I} (\cos(\varphi) + \\
    & \cos(2 \omega t + 2 \phi_v) \cos(\varphi) + \sin(2 \omega t + 2 \phi_v) \sin(\varphi)) \\
    &= P(1 + \cos(2 \omega t + 2 \phi_v)) + Q \sin(2 \omega t + 2 \phi_v)
\end{aligned}
\end{equation}
where $\varphi = \phi_v - \phi_i$, $P = \bar{V} \bar{I} \cos(\varphi)$ is called active power and $Q = \bar{V} \bar{I} \sin(\varphi)$ is called reactive power.
The amount of power consumed by a circuit element on average per period ($T$) is equal to $P$:
\begin{equation}
\label{eq:average-power}
    \frac1T \int_0^T p(t) dt = P
\end{equation}

\subsection{Phasors}

In many cases, periodic oscillations of physical quantities, such as voltage or current, do not matter because the relevant behavior can instead be described by period-averaged quantities, as in Eq.~(\ref{eq:average-power}).
In these cases, it is convenient to use \textit{phasors} notation.
A phasor is a complex number whose amplitude corresponds to the RMS value of the represented quantity and angle corresponds to its phase.
For example, current phasor representing Eq.~(\ref{eq:current}) is given by
\begin{equation}
    I = \bar{I} e^{i \phi_i}
\end{equation}
A time-dependent representation of the current can be recovered from its phasor by
\begin{equation}
\label{eq:time-phasor-conversion}
    i(t) = \sqrt{2} Re(I e^{i \omega t})
\end{equation}
Voltage phasor is also defined similarly:
\begin{equation}
    V = \bar{V} e^{i \phi_v}
\end{equation}

Voltage and current phasors are related through a generalization of Ohm's law: $V = I Z$, where $Z$ is a complex number called \textit{impedance}, which is a generalization of resistance.
Impedance is defined as $Z = R + iX$, where $R$ is resistance and $X$ is reactance.
An inverse property $Y = 1 / Z$ is called \textit{admittance}.

Similarly to Eq.~(\ref{eq:power}), a product of voltage and (complex-conjugated) current phasors gives us a quantity $S$, called \textit{complex power}.
\begin{equation}
\label{eq:complex-power}
    S = V I^* = \bar{V} \bar{I} e^{i \varphi} = P + i Q
\end{equation}
Note that $S$ is not a phasor of power per se, since we cannot recover time-dependent $p(t)$ from it using an equivalent of Eq.~(\ref{eq:time-phasor-conversion}), but it still gives us information about the values of $P$ and $Q$, which is sufficient for many purposes.

\subsection{Distributed Power Networks}

Real-life electrical networks are complicated structures connecting together a large number of power consumers and sources (generators).
At some level of abstraction, they can be represented as a graph where nodes represent so-called \textit{buses}.
A bus is a junction point in the network where multiple consumers and generators can be connected.
The edges between nodes represent electrical \textit{lines}, typically long-range conductors, connecting different buses into a single network.

Electricity typically requires closed current paths.
In real networks these paths also involve additional conductors (e.g. other phase conductors) together with the source and load connections.
These additional conductors are not explicitly included in the graph abstraction since it is assumed that their electrical quantities can be inferred from the model variables, so they introduce no additional degrees of freedom.
For example, in a typical 3-phase system under normal circumstances voltages and currents in each of the 3 phase conductors have the same amplitude, frequency, and only differ by phase (0 and $\pm 2 \pi / 3$).

Local connections between a power station (where buses are located) and final consumers or generators are assumed to be ideal and are also not explicitly included in the graph model.
Thus, graph nodes can also be thought of as representing neighborhoods powered by a particular power station.

Individual consumers are typically not modeled explicitly either. 
Instead a total cumulative load from all consumers at each node is considered (a single constant complex number).
Of course, individual power consumers are typically not constant in time. 
However if the number of consumers is sufficiently large it is usually a good approximation to treat the cumulative load as constant.

On the other hand, the number of power sources (generators) is typically much smaller than the number of consumers, so each individual generator is modeled explicitly.

The continuous variables that need to be optimized in the network model are the voltage phasors at each bus $V_i$ (where voltage magnitudes are defined with respect to a common implicit reference) and complex power outputs of each generator $S_g$. 
Because of the network equations and constraints (see Eqs.~\eqref{eq:node-balance}--\eqref{eq:first-constr} below) these variables are not independent, and typically cannot be controlled directly in real systems, so they should not be regarded as primary control knobs of the system.
Instead, they should be treated as a consequence of the primary controls (typically electromotive force), whose optimized values can be used to recover the optimal settings of the system.

\section{Theory}
\label{sec:theory}

\subsection{Power Network Model}

In this work we consider a single-period AC-OPF-UC problem.
Typical UC intertemporal constraints, such as ramping and minimum up/down times, are not considered here for the purpose of isolating the binary commitment component, while keeping the benchmark tractable.

\begin{figure}
    \centering
    \includegraphics[width=\linewidth]{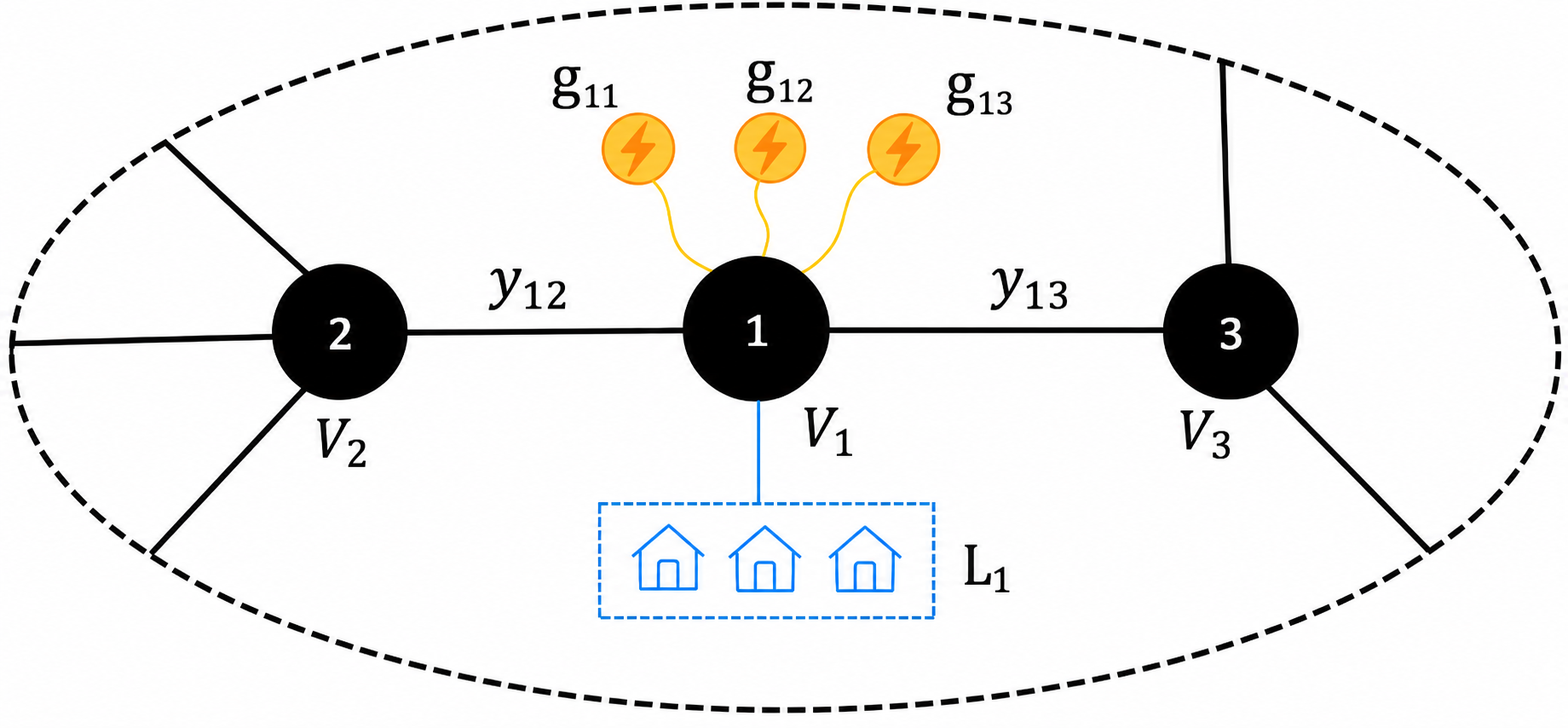}
    \caption{A section of an input graph. The nodes represent buses and have associated voltage phasors ($V$), cumulative loads ($L$), and connected generators ($g$). The edges represent electrical lines and have associated admittances ($y$).}
    \label{fig:graph_example}
\end{figure}

As an input we get a graph describing configuration and parameters of a given power network (Figure~\ref{fig:graph_example}). 
Each node $i$ of the graph has its associated voltage phasor ($V_i = v_ie^{i\delta_i}$) and specifies the minimum and maximum values of voltage magnitudes ($v_i^{min}, v_i^{max}$), as well as its cumulative power load from all consumers $L_i$ (complex) and a list of connected generators $G_i = \{g_{ij}\}$.
Each edge specifies its admittance ($Y_{ij}$) and maximum supported current ($|I_{ij}|^{max}$).

Each generator $g \in G_i$ (subscripts on $g$ are implicit for less cluttered notation) has an associated power output $S_g = P_g + i Q_g$, status $u_g$ (off=0 or on=1) and specifies its minimum and maximum values of active and reactive powers ($P_g^{min}, P_g^{max}, Q_g^{min}, Q_g^{max}$) and cost terms ($a_g, b_g, c_g$).
The cost of producing a given level of power output ($C_g$) is given by 
\begin{equation}
\label{eq:power-cost}
    C_g(u_g, P_g) = u_g(a_g P^2_g + b_g P_g + c_g)
\end{equation}
The reason $Q$ is not a part of the cost is because it affects the instantaneous amplitude of power, but does not contribute to net power consumption per period, see Eq~(\ref{eq:average-power}).
Note that $c_g$ is a flat cost of turning the generator on, which does not depend on the amount of generated power.

The net power at node $i$ is the total power produced by all generators connected to the node minus the power consumed by consumers and plus/minus power flow from/to other nodes, i.e.
\begin{equation}
\label{eq:node-balance}
    S_i = \sum_{g \in G_i} u_g S_g - L_i - \sum_{j \in N(i)} S_{ij}
\end{equation}
where $N(i)$ is a set of neighbors of node $i$ and $S_{ij} = P_{ij} + iQ_{ij}$ is the complex power flow from node $i$ to node $j$. 
Negative values of $P_{ij}$ or $Q_{ij}$ mean that the corresponding power (active or reactive) is flowing in the opposite direction (from $j$ to $i$). 
The power leaving node $i$ toward node $j$ ($S_{ij}$) is given by 
\begin{equation}
    S_{ij} = V_i I_{ij}^*
\end{equation}
where $I_{ij}$ is the current phasor for line $i \rightarrow j$, and given by
\begin{equation}
    I_{ij} = Y_{ij} (V_i - V_j)
\end{equation}

Note that the line itself always absorbs active power, i.e.
\begin{equation}
    P_{ij} + P_{ji} = R_{ij} |I_{ij}|^2
\end{equation}
Assuming a power flow from $i$ to $j$, active power received by node $j$ ($-P_{ji}$) will always be less than the active power sent by node $i$ ($P_{ij}$), since line resistance $R_{ij} > 0$. 
Reactive power received by node $j$ can be more or less than the reactive power sent by node $i$ depending on the sign of line reactance ($X_{ij}$), which could be both positive (inductive behavior, more common) or negative (capacitive behavior).

The goal of the problem is to find the values of the variables ($V_i$, $S_g$ and $u_g$) such that the total cost ($C$) of power production due to all generators in the network and voltage magnitude deviation from nominal (=1) due to all nodes in the network is minimized
\begin{equation}
\label{eq:cost}
    C = \sum_g C_g + \lambda \sum_i (v_i - 1)^2
\end{equation}
the nodal power balance is satisfied,
\begin{equation}
\label{eq:first-constr}
    S_i = 0
\end{equation}
RMS current through each line is within its capacity,
\begin{equation}
\label{eq:capacity}
    |I_{ij}| \le |I_{ij}|^{max}
\end{equation}
and the limiting values of the variables are respected.
\begin{gather}
    \label{eq:p-limits}
    P_g^{min} \le P_g \le P_g^{max} \\
    \label{eq:q-limits}
    Q_g^{min} \le Q_g \le Q_g^{max} \\
    \label{eq:last-constr}
    v_i^{min} \le v_i \le v_i^{max}
\end{gather}
Here, $\lambda$ is an arbitrary weight factor balancing the cost of voltage deviation objective relative to power production cost.

The voltage phases $\delta_i$ only make sense relative to one another, so to remove the global phase degeneracy we fix the phase of node 0 (chosen arbitrarily) to 0, which removes one of the model variables ($\delta_0$).

To summarize, there is a total of $2N - 1 + 2|G|$ real continuous variables and $|G|$ binary variables to optimize, where $N$ is the total number of nodes, and $|G|$ is the total number of generators across all nodes. 
The total number of equality constraints in real variables is $2N$ and the total number of inequality constraints is $2N + 4|G| + E$, where $E$ is the total number of edges in the graph.

\subsection{Constraints in Model Variables}
\label{sec:real-variables}

Eqs.~\eqref{eq:node-balance} and \eqref{eq:capacity} are specified in terms of $S_i$ and $|I_{ij}|$, which can be expressed in terms of model variables, but are not model variables themselves.
In this section we derive the expressions for those constraints in terms of model variables: ($u_g, P_g, Q_g, v_i, \delta_i$).

First, let us derive the equations for the real and imaginary parts of line power flows $S_{ij}$:
\begin{equation}
    \begin{aligned}
    \label{eq:sij}
        S_{ij} &= P_{ij} + i Q_{ij} = V_i I_{ij}^* = Y_{ij}^* (|V_i|^2 - V_i V_j^*) \\
        &= (\alpha_{ij} - i \beta_{ij}) (v_i^2 - v_iv_j (\cos{\delta_{ij}} + i \sin{\delta_{ij}}))
    \end{aligned}
\end{equation}
where $Y_{ij} = \alpha_{ij} + i \beta_{ij}$ and $\delta_{ij} = \delta_i - \delta_j$.

Therefore,
\begin{equation}
\label{eq:pij}
    P_{ij} = \alpha_{ij} v_i^2 - v_i v_j (\alpha_{ij} \cos \delta_{ij} + \beta_{ij} \sin \delta_{ij})
\end{equation}
\begin{equation}
\label{eq:qij}
    Q_{ij} = -\beta_{ij} v_i^2 + v_i v_j (\beta_{ij} \cos \delta_{ij} - \alpha_{ij} \sin \delta_{ij})
\end{equation}
and $S_i = 0$ constraint translates to
\begin{equation}
    Re(S_i) = \sum_{g \in G_i} u_g P_g - Re(L_i) - \sum_{j \in N(i)} P_{ij} = 0
\end{equation}
\begin{equation}
    Im(S_i) = \sum_{g \in G_i} u_g Q_g - Im(L_i) - \sum_{j \in N(i)} Q_{ij} = 0
\end{equation}
where the values of $L_i$ are given constants.

Squared magnitude of current phasor in terms of model variables is
\begin{equation}
    \begin{aligned}
    \label{eq:iij}
        |I_{ij}|^2 &= |Y_{ij}|^2 |V_i - V_j|^2 \\
        & = |Y_{ij}|^2 \left( v_i^2 + v_j^2 - 2 v_i v_j \cos(\delta_{ij}) \right)
    \end{aligned}
\end{equation}

Then the capacity constraint (Eq.~\eqref{eq:capacity}) can be equivalently written as
\begin{equation}
    |I_{ij}|^2 \leq (|I_{ij}|^{max})^2
\end{equation}

\subsection{Full Encoding Algorithm}
\label{sec:discretized}

Variational quantum algorithms (VQAs), such as Quantum Approximate Optimization Algorithm (QAOA), have been commonly used for other power grid problems
\cite{ajagekar2019quantum, koretsky2021adapting, salgado2024hybrid, magar2024dc, wang2026review}.
As such, we are first going to consider a straightforward application of QAOA for this problem with full encoding of the cost function on the quantum computer.

In short, QAOA applies alternating layers of two unitary operators, the \textit{mixing operator} and the \textit{cost operator}, to an initial quantum state.
In the standard construction, used here, the initial state is the uniform superposition over all computational-basis bitstrings.
More structured initial states can bias the search distribution \cite{egger2021warm, tate2023warm, he2023alignment}, but require additional state-preparation or circuit-synthesis primitives \cite{grover2002creating, mottonen2004transformation, shende2005synthesis, gonzales2025efficient, atallah2026simulating, gaidai2026decomposition}.

Mixing operator ($\hat{B}$) is problem-independent and is typically chosen as
\begin{equation}
    \label{eq:mixer}
    \hat{B} = \sum_i \hat{X_i}
\end{equation}
where $\hat{X_i}$ is Pauli-X operator.
Another common choice is constraint-preserving mixers \cite{hadfield2019quantum, fuchs2022constraint, bartschi2020grover, wilkie2025learning}, but they are challenging to design for non-trivial constraints, therefore we will stick to the default choice of Eq.~(\ref{eq:mixer}).

Since constraints are not enforced by the mixer, they will need to be added as penalty terms to the cost function. 
Therefore, in this approach our total (penalized) cost function becomes
\begin{equation}
\label{eq:penalized-cost}
    \tilde{C} = C + \lambda_2 \sum_k d_k^2
\end{equation}
where $C$ is defined in Eq.~(\ref{eq:cost}), $d_k$ is the left-hand side of an equality constraint of the form $d_k = 0$, the sum is over all model constraints, and $\lambda_2$ is a variable factor that adjusts the cost of constraint violation with respect to the cost of the primary objective value.

Inequality constraints are converted to equality constraints via the slack variable approach \cite{nocedal2006numerical}.
Namely, each inequality constraint of the form $d_k \le 0$ is written as an equivalent equality constraint $d_k + s_k = 0$, where a new variable $s_k \ge 0$ is introduced, called a \textit{slack variable}.

The cost operator for QAOA is usually a unitary operator corresponding (via exponentiation) to a Hamiltonian ($\hat{H}_C$) such that
\begin{equation}
    \label{eq:cost-hamiltonian}
    \hat{H}_C \ket{x} = \tilde{C}(x) \ket{x}
\end{equation}
where $x$ is a bitstring encoding the values of all model variables, although alternatives exist as well \cite{wilkie2024quantum}.

Binary variables can be encoded straightforwardly -- they map to the qubits directly.
Continuous variables will be encoded on a discrete grid of values.
Specifically, each generic continuous variable $y \in \{P, Q, v, \delta, s\}$ is represented on a grid starting from $y_{min}$ and with a fixed step size of $\Delta_y$ as
\begin{equation}
\label{eq:binary-discretization}
    y = y_{min} + \Delta_y \sum_{k = 0}^{n_y - 1} 2^k y_{k}
\end{equation}
where $y_{k} \in \{0,1\}$ is the value of $k$-th bit in the binary encoding of $y$ and $n_y$ is the total number of bits used for the encoding.
Using higher values of $n_y$ allows more precise continuous variable encoding, but also requires more qubits.

Note that non-negativity of slack variables is automatically enforced by defining non-negative grids for them.
Similarly, simple box constraints, such as Eqs.~\eqref{eq:p-limits} and \eqref{eq:q-limits}, can also be enforced by defining their grids within their respective bounds, so these inequalities do not need slack variables.

To avoid trigonometric terms in the constraints (see Sec.~(\ref{sec:real-variables})), voltage phasor can be parametrized in rectangular coordinates instead:
\begin{equation}
\label{eq:voltage-rectangular}
    V_i = V_{R, i} + i V_{I, i}
\end{equation}

Because of the new parametrization, the following equations will change. 
Voltage magnitude constraint (Eq.~\eqref{eq:last-constr}) stops being a simple box constraint and instead becomes
\begin{equation}
    (v_i^{min})^2 \le V_{R, i}^2 + V_{I, i}^2 \le (v_i^{max})^2
\end{equation}
Complex power flow equation (Eq.~\eqref{eq:sij}) becomes
\begin{equation}
    \begin{aligned}
        S_{ij} &= Y_{ij}^* (|V_i|^2 - V_i V_j^*) \\
        &= (\alpha_{ij} - i \beta_{ij}) (V_{R, i}^2 + V_{I, i}^2 - V_{R, i} V_{R, j} - V_{I, i} V_{I, j} \\
        &+ i(V_{R, i} V_{I, j} - V_{I, i} V_{R, j}))
    \end{aligned}
\end{equation}
Then real and imaginary flow equations (Eqs.~\eqref{eq:pij} and \eqref{eq:qij}) become
\begin{equation}
    \begin{aligned}
        P_{ij} &= Re(S_{ij}) \\
        &= \alpha_{ij} (V_{R, i}^2 + V_{I, i}^2 - V_{R, i} V_{R, j} - V_{I, i} V_{I, j}) \\
        &+ \beta_{ij} (V_{R, i} V_{I, j} - V_{I, i} V_{R, j})
    \end{aligned}
\end{equation}
\begin{equation}
    \begin{aligned}
        Q_{ij} &= Im(S_{ij}) \\
        &= \alpha_{ij} (V_{R, i} V_{I, j} - V_{I, i} V_{R, j}) \\
        &- \beta_{ij} (V_{R, i}^2 + V_{I, i}^2 - V_{R, i} V_{R, j} - V_{I, i} V_{I, j})
    \end{aligned}
\end{equation}
Current phasor magnitude expression (Eq.~\eqref{eq:iij}) becomes
\begin{equation}
    \begin{aligned}
    \label{eq:current-magnitude-rectangle}
        |I_{ij}|^2 &= |Y_{ij}|^2 |V_i - V_j|^2 \\
        &= |Y_{ij}|^2 \left( (V_{R, i} - V_{R, j})^2 + (V_{I, i} - V_{I, j})^2 \right)
    \end{aligned}
\end{equation}
Finally, instead of setting $\delta_0 = 0$ we set $V_{I, 0} = 0$ and $V_{R, 0} \ge 0$. 

Plugging Eq.~\eqref{eq:binary-discretization}-\eqref{eq:current-magnitude-rectangle} into Eq.~\eqref{eq:penalized-cost}, one can obtain the cost function in terms of individual binary variables $x_{k}$ (from Eq.~\eqref{eq:cost-hamiltonian}). Then the cost Hamiltonian can be obtained by the usual substitution
\begin{equation}
    x_{k} \rightarrow \frac{1 - \hat{Z}_{k}}{2}
\end{equation}
where $\hat{Z}_{k}$ is the Pauli-Z operator acting on $k$-th qubit.

The continuous variable grid in Eq.~\eqref{eq:binary-discretization} can be sequentially refined. 
Specifically, once a set of optimal values on their corresponding grids is found in a particular run, one can redefine the grids to make them denser around the optimal values, or shift the grid if optimum is on the edge, thus obtaining more precise values in the next run.
This way the number of qubits necessary to achieve a given precision level in continuous variables can be reduced at the cost of larger number of runs.
Of course, if the initial grid is too coarse it is possible to miss the global minimum entirely, so some parameter tuning might be necessary.

Accounting for slack variables, the total number of continuous model variables becomes $4N - 1 + 2|G| + E$.
Then the total number of qubits required, assuming constant $n_y$, is $n_y (4N - 1 + 2|G| + E) + |G|$.

As an example, assuming average node degree = 3 (so $E = 3/2 N$), $N = |G|$ and $n_y = 2$ we get $\approx\!\!16|G|$ qubits, so even a small 10 generator problem would require 160 qubits to fully encode the problem even at this minimal discretization level.

\subsection{Qubit-Efficient Algorithm}
\label{sec:qubit-efficient-algorithm}

The direct application of QAOA to AC-OPF-UC considered in the previous section requires a very large number of qubits even for small instances.
In order to reduce qubit requirements, one could encode only a part of the problem variables on a quantum computer, instead of encoding all of them as in Section~\ref{sec:discretized}.
The primary candidates for the encoding are the binary variables ($u_g$), since they are harder to optimize classically and map naturally to qubits.
Encoding only binary variables requires only $|G|$ qubits, which is a dramatic decrease compared to the $\approx\!\!16|G|$ result of Section \ref{sec:discretized}.
The main issue with this approach is that without encoding other variables it is no longer possible to evaluate the cost function directly on a quantum computer, so QAOA cannot be used.

Several recent studies \cite{aboumrad2025new, xie2025predict, azzam2014mixed} of related problems get around this issue by considering surrogate cost functions, which are simpler to evaluate and optimize functions that typically do not depend on all variables and/or depend on them in a simplified way, but are expected to have their minimum in the vicinity of the true cost function's minimum. 
Ideally, coinciding with it.
If such function is available, then finding the minimum of it, together with a simple local search around it, could reveal the true cost's minimum.
However, design of such functions is a non-trivial problem on its own, and they often do not work in full generality, suitable only for a particular subspace of parameters.

\begin{figure}
    \centering
    \includegraphics[width=\linewidth]{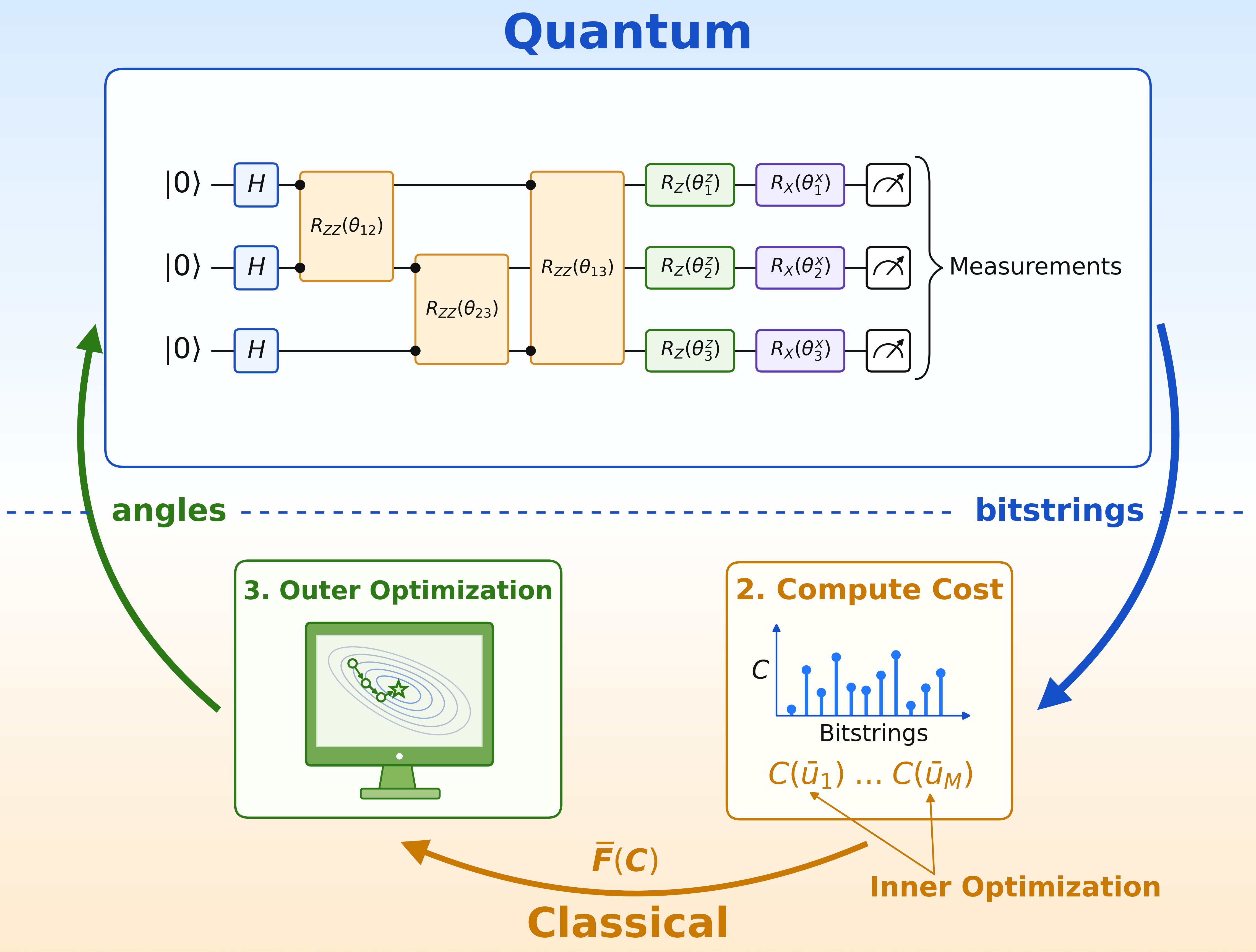}
    \caption{Schematic representation of the workflow of the algorithm described in Section \ref{sec:qubit-efficient-algorithm}. Bitstrings ($\bar{u}_1$, ..., $\bar{u}_M$) are measured from the quantum circuit. Their costs are evaluated classically by optimizing continuous degrees of freedom. Then an average target metric $\bar{F}(C)$ is passed to the outer optimization layer over variational angles. The new variational angles proposed by the optimizer are applied to the quantum circuit. See the text for more details.}
    \label{fig:workflow-outline}
\end{figure}

Instead, we will use a more general approach in which the cost function is still exact, but is evaluated classically. Specifically, every time a particular bitstring $\bar{u}_i = u_1 ... u_{|G|}$ is sampled from a quantum computer, its cost $C_i = C(\bar{u}_i)$ can be evaluated by running a classical optimization over remaining continuous variables, while keeping generator statuses fixed to the values of $u_1 ... u_{|G|}$ (\textit{inner} optimization).
From the cost we can calculate the target metric $F(C_i)$, which can be the cost itself or another map such as Approximation Ratio (AR).
Repeating this for many samples, we can calculate sample mean value of the target metric $\bar{F}(C)$ for the current set of variational angles -- an estimator for its expectation -- and return that value to the \textit{outer} optimization layer, which looks for variational angles that optimize the target metric expectation.
The new angles suggested by the optimizer, which include angles for all gates ($R_{ZZ}$, $R_Z$, $R_X$) are then applied to the quantum circuit and the process repeats.
This approach is shown schematically in Figure~\ref{fig:workflow-outline}.

As we sample bitstrings during the variational angle optimization, we keep track of the best bitstring sampled so far (i.e. lowest cost or highest AR bitstring) and return it as the answer encoding best-known generator assignments, once the variational angles optimization is finished.
Alternatively, we could keep sampling from the optimized probability distribution or start optimization again from different initial angles until either a given quality threshold is surpassed or we run out of the optimization time budget.

Since the cost is evaluated classically now, we can evaluate non-polynomial functions, so we no longer need to resort to rectangular voltage coordinates of Eq.~\eqref{eq:voltage-rectangular} and can use the equations of Sec.~(\ref{sec:real-variables}).
We also do not need to introduce slack variables to convert inequality constraints to equality.
Instead the corresponding constraint violation term $d_k$ of Eq.~\eqref{eq:penalized-cost} can be defined directly as
\begin{equation}
    d_k = \mathrm{max}(\tilde{d}_k, 0)
\end{equation}
for an inequality constraint of the form $\tilde{d}_k \le 0$.

\section{Results}
\label{sec:results}

The algorithm presented in Section~(\ref{sec:discretized}) has too high qubit requirement to simulate it classically, so in this section we are going to focus on the qubit-efficient algorithm from Section~(\ref{sec:qubit-efficient-algorithm}).

The VQA ansatz that we used in this benchmark consists of 1 layer made up of single-qubit Hadamard gates, followed by $R_{ZZ}$ gates between all pairs of qubits, followed by $R_Z$ and $R_X$ gates on all qubits (see Figure~\ref{fig:workflow-outline}).
Thus, the total number of variational angles in our ansatz is equal to $Q (Q - 1) / 2 + 2Q = 0.5Q^2 + 1.5Q$ where $Q$ is the number of qubits in the circuit.
For our binary encoding $Q = |G|$.

\subsection{Datasets}
\label{sec:datasets}

To benchmark the performance of the algorithm we created 9 datasets with number of generators ranging from 5 to 13, and 120 random problem instances in each set.
The underlying graph for each problem was generated as a random geometric graph on a unit square with number of nodes equal to the number of generators ($N = |G|$) and connection radius chosen such that the expected average node degree is equal to 3.
The $N = |G|$ constraint was chosen as the simplest fixed ratio between nodes and generators, since it is naturally expected that these parameters would be co-dependent.
Disconnected components were made connected by adding an edge between their closest nodes.

Many of the values in this benchmark were drawn from lognormal distributions.
A random variable $X$ is said to be distributed according to a lognormal distribution if $\log(X)$ is distributed according to a normal distribution:
\begin{equation}
    X \sim \mathrm{LogNormal}(\mu, \sigma^2) \leftrightarrow
    \log(X) \sim \mathrm{Normal(\mu, \sigma^2)}
\end{equation}
i.e. $\log(X)$ has mean $\mu$ and standard deviation $\sigma$.

Instead of parameterizing our lognormal distributions by $\mu$ and $\sigma$ directly, we parameterize them by the mean value of $X$ (=$m$) and its spread factor, defined as a number $F$ such that 90\% of the distribution is contained within the range of $[m / F, mF]$, i.e.
\begin{equation}
    P\left( \frac{m}{F} \le X \le mF \right) = 0.9
\end{equation}

The value of $\sigma$ corresponding to a given value of $F$ can be found numerically by solving
\begin{equation}
    \Phi\left( \frac{\sigma}{2} + \frac{\log(F)}{\sigma} \right) - \Phi\left( \frac{\sigma}{2} - \frac{\log(F)}{\sigma} \right) = 0.9
\end{equation}
where $\Phi$ is the standard normal cumulative distribution function (CDF).
The value of $\mu$ can then be calculated as
\begin{equation}
    \mu = \log(m) - \frac{\sigma^2}{2}
\end{equation}

The absolute value of the complex load at each node in the graph ($|L_i|$) was sampled from a lognormal distribution with $m = 1$ and $F = 10$.
The large values of spread used here and further below were chosen deliberately to test a broad range of instances rather than reproduce typical grid parameters.
Reactive fraction of the load ($f_i^L$) at each node was sampled from a uniform distribution in the [0, 0.1] range.
Then the full complex load was calculated as
\begin{equation}
    L_i = |L_i| \sqrt{1 - (f_i^L)^2} + i |L_i| f_i^L
\end{equation}

The voltage limits for each node were defined as [0, 100].
Note that the voltage deviations from nominal (=1) are explicitly penalized by the cost function of Eq.~\eqref{eq:cost}, so these wide bounds are only exercised as a last resort when instance feasibility cannot be achieved otherwise.

Each edge's impedance was generated similarly.
The absolute value ($|Z_{ij}|$) was initially sampled from a lognormal distribution with $m = 0.01$ and $F = 10$.
Then the sampled value was additionally multiplied by a ratio of the edge length (which is well-defined for geometric graphs) to median edge length to give larger impedance to longer edges and vice versa.
Reactive fraction of the impedance ($f_{ij}^X$) was sampled from a uniform distribution in the [0.9, 1] range.
Then the full edge impedance was calculated as
\begin{equation}
    Z_{ij} = |Z_{ij}| \sqrt{1 - (f_{ij}^X)^2} + i |Z_{ij}| f_{ij}^X
\end{equation}
and admittance as $Y_{ij} = 1 / Z_{ij}$.

The current capacity of each edge was sampled from a lognormal distribution with $m = 8 \frac{m(|L|)}{m(d)} = 8 / 3$ and $F = 2$.
Here and further $m(\cdot)$ denotes the mean value of its argument, and $d$ is a node degree.
The ratio of $m(|L|) / m(d)$ gives us expected load per line, which is approximately equal to expected current per line, assuming voltage is close to 1.
The factor of 8 just adds some room for variance on top of that.

Generators were assigned to nodes randomly according to a degree-biased multinomial distribution.
Specifically, the probability of assigning a generator to node $i$ was defined as
\begin{equation}
    p_i = \frac{w_i}{\sum_j w_j}
\end{equation}
where
\begin{equation}
    w_i = \mathrm{exp}\left( \frac{\beta}{1 - \beta} (d_i - d_{\mathrm{max}}) \right)
\end{equation}
where $d_i$ is node degree, $d_{\mathrm{max}}$ is the maximum node degree in the graph, and $\beta \in [0, 1]$ is a degree-bias parameter.
When $\beta = 0$, all nodes have equal probabilities of getting a generator.
When $\beta = 1$, only maximum degree nodes have non-zero probability of getting a generator.
For this benchmark $\beta = 0.3$ was chosen.

The range of absolute value of power for each generator was parameterized via its multiplicative midpoint ($S^{ref}_g$) and length factor ($LF_g$).
$S^{ref}_g$ was sampled from a lognormal distribution with $m = 1.2 m(|L|) = 1.2$ and $F = F(|L|) = 10$, where $F(|L|)$ is the value of $F$ for the distribution of absolute value of a node's load.
The length factor was calculated as $LF_g = 1 + X_g$, where $X_g$ was sampled from a lognormal distribution with $m = 0.5$ and $F = 1.2$.
The reference values of active and reactive powers were calculated as
\begin{equation}
    P_g^{ref} = S_g^{ref} \sqrt{1 - (f_g^G)^2}
\end{equation}
\begin{equation}
    Q_g^{ref} = S_g^{ref} f_g^G
\end{equation}
where $f_g^G$ is generators reactive fraction sampled from a uniform distribution in the [0, 0.1] range.
Then the minimum and maximum values of active power were calculated as
\begin{equation}
    P_g^{min} = P_g^{ref} / LF_g, \quad P_g^{max} = P_g^{ref} LF_g
\end{equation}
And the minimum and maximum values of reactive power were calculated as
\begin{equation}
    Q_g^{min} = -Q_g^{ref} LF_g, \quad Q_g^{max} = Q_g^{ref} LF_g
\end{equation}

The cost coefficients $a, b$ and $c$ for each generator were sampled from their respective lognormal distributions with means $m_a = 1 / m(P^{ref})^2$, $m_b = 1 / m(P^{ref})$ and $m_c = 1$; and spreads $F_a = 2$, $F_b = 1.5$, and $F_c = 2$.

After an instance is generated, a fast aggregate check was also run on it to filter out bad instances (e.g. apparently close to infeasible).
An instance was deemed bad if:

1) Total maximum active and reactive generator power is less than total active or reactive load (or close to it), i.e.
\begin{equation}
    \sum_g P_g^{max} < \gamma \sum_i Re(L_i)
\end{equation}
or
\begin{equation}
    \sum_g Q_g^{max} < \gamma \sum_i Im(L_i)
\end{equation}
where $\gamma$ is an arbitrary strictness factor controlling probability of infeasibility. 
If this inequality is true with $\gamma = 1$ then the instance is definitely infeasible.
As the value of $\gamma$ is increased beyond 1, infeasibility probability is getting smaller.
In this benchmark the value of $\gamma$ was chosen to be 1.2 in this and the following checks.

2) Some generators cannot be turned on because their minimum active or reactive power output is already greater than total active or reactive load (or close to it), i.e.
\begin{equation}
    \max_g P_g^{min} > \gamma^{-1} \sum_i Re(L_i)
\end{equation}
or
\begin{equation}
    \max_g Q_g^{min} > \gamma^{-1} \sum_i Im(L_i)
\end{equation}

3) Minimum necessary current that needs to be imported to any node $i$ exceeds total incident edge capacity (or close to it), i.e.
\begin{equation}
    \sum_{j \in N(i)} |I_{ij}|^{max} < \gamma |I_i|^{imp}
\end{equation}
where
\begin{equation}
    |I_i|^{imp} = \frac{\sqrt{(P_i^{imp})^2 + (Q_i^{imp})^2}}{v_i^{max}}
\end{equation}
and
\begin{equation}
    P_i^{imp} = \max\left( 0, Re(L_i) - \sum_{g \in G_i} P_g^{max} \right)
\end{equation}
\begin{equation}
    Q_i^{imp} = \max\left( 0, Im(L_i) - \sum_{g \in G_i} Q_g^{max} \right)
\end{equation}

Note that while these checks only filter out obviously infeasible instances and do not guarantee feasibility, in practice they were sufficient to completely remove all infeasible instances from our datasets.

\subsection{Target Metric}

The target metric that we optimize in this benchmark is approximation ratio (AR).
AR of a given bitstring $i$ is defined as a ratio of the best-known bitstring cost to the cost of bitstring $i$, i.e.
\begin{equation}
    AR_i = C_{best} / \tilde{C}_i
\end{equation}
where $C_{best}$ is the lowest cost found by explicitly optimizing all $2^{|G|}$ bitstrings, and $\tilde{C}_i$ is the full constraint-penalized cost calculated from Eq.~\eqref{eq:penalized-cost} with $\lambda_2 = 10^7$, which was found experimentally as the smallest value that ensures that all infeasible bitstrings have higher total costs than the best-known feasible one.

Since IPOPT is a local solver for a nonconvex problem, $C_{best}$ is not a certified global optimum.
Consequently, AR $=1$ corresponds to the best solution found in our benchmark.

The values of AR are in [0, 1] range for any bitstring of any instance (larger = better), which allows us to calculate meaningful instance-average AR as a metric for the entire dataset, and also avoid numerical issues due to very large penalty costs on some of the infeasible bitstrings.

When optimization begins, the value of $C_{best}$ is not yet known.
However, for the purpose of optimization one can equivalently an arbitrary constant (e.g. a lower bound on $C_{best}$) instead of its actual value.
This will change the range of attainable AR values, but it can be rescaled later and it does not invalidate the mentioned advantages of using AR over plain cost.

\subsection{Number of Shots}

The first question that we want to answer is how many shots are required to reliably estimate the expectation of AR for a given set of variational angles.
The following approach was used to answer that question.

For a given instance and a given set of variational angles, we can calculate the exact probability distribution over all bitstrings of the instance and from that calculate exact expectation ($\mu$) and standard deviation ($\sigma$) of their AR values.
Then we can use that information to calculate the minimum number of shots ($M$) such that sample mean over $M$ shots drawn from that distribution ($\overline{AR}$) falls within the range of $\mu$ $\pm$ target confidence interval half-length ($r$) with target probability ($\alpha$).
In other words, the goal is to find $M$ such that $\overline{AR}(M)$ satisfies
\begin{equation}
    P(|\overline{AR} - \mu| \le r) \ge \alpha
\end{equation}
Using Bernstein's inequality \cite{boucheron2013concentration}, one can calculate $M$ value as
\begin{equation}
\label{eq:num-shots}
    M = \left\lceil \frac{(2\sigma^2 + 2r/3) \log(2 / (1 - \alpha))}{r^2} \right\rceil
\end{equation}
In this benchmark we used $r = 0.05$ and $\alpha = 0.9$.
Since $AR\in[0,1]$, its variance satisfies $\sigma^2\le1/4$. Therefore, Eq.~\eqref{eq:num-shots} gives a theoretical worst-case bound of $M=640$ shots for the chosen $r$ and $\alpha$.

\begin{figure}
    \centering
    \includegraphics[width=\linewidth]{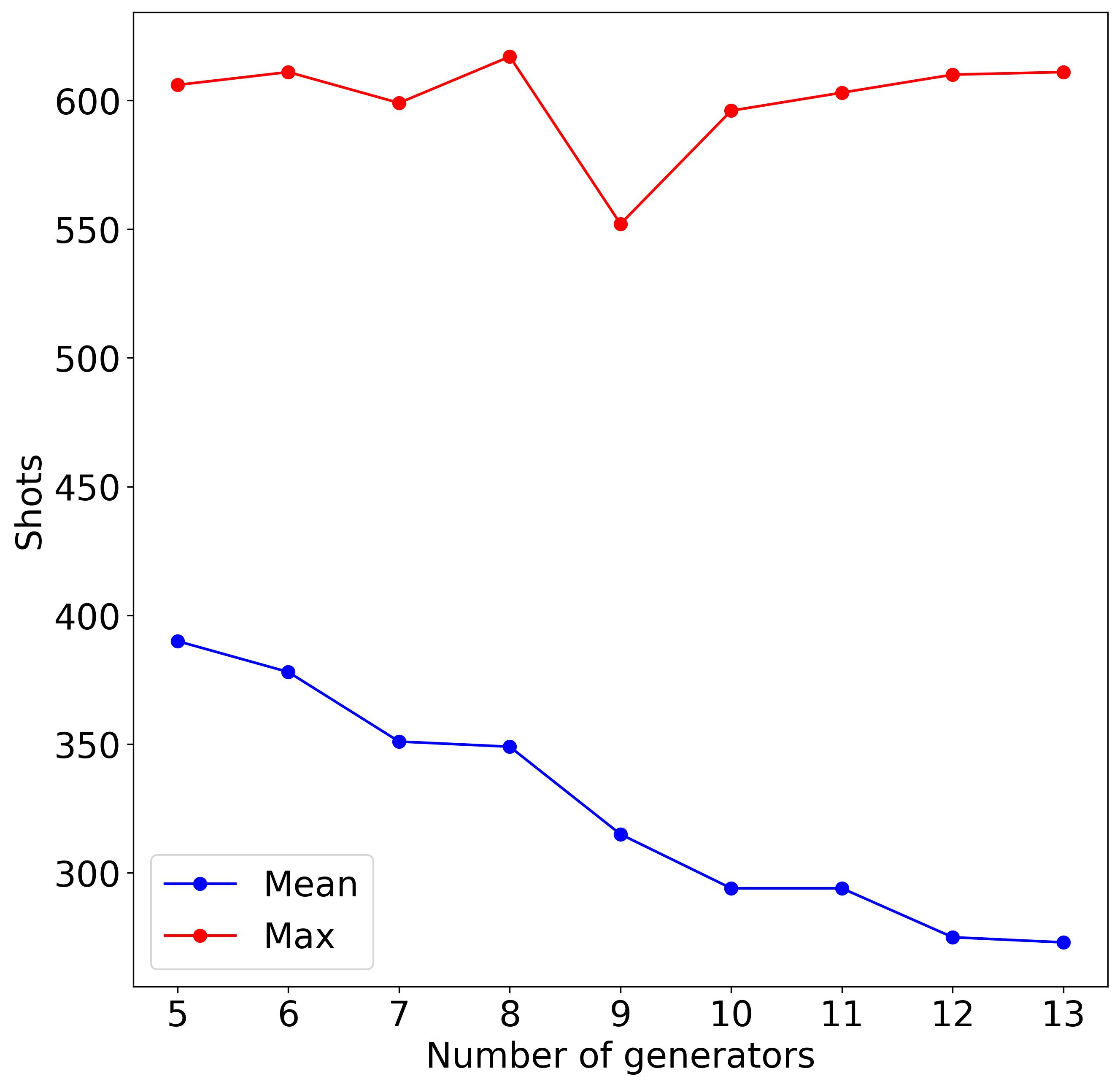}
    \caption{Mean and maximum number of shots necessary to converge mean sample AR to within 0.05 of its expected value with probability 90\%. The mean and maximum is taken over different random variational angles and random instances at a given system size $N = |G|$.}
    \label{fig:shots-vs-gens}
\end{figure}

For each instance in each dataset, we generated 100 random angle sets, used Eq.~\eqref{eq:num-shots} to calculate their values of $M$, and plotted the mean and maximum of those values across all instances and all angle sets of a given dataset in Figure~\ref{fig:shots-vs-gens}.

These results have also been verified with Monte-Carlo simulation.
Specifically, for each instance and each angle set, we sampled $M$ values from the corresponding probability distribution, as calculated by Eq.~\eqref{eq:num-shots}, and calculated sample mean ($\overline{AR}$).
This was repeated 10000 times.
Then we calculated the fraction of samples that fell into the target range of $\mu \pm r$.
Treating each sample as a Bernoulli trial with theoretical success probability equal to $\alpha = 0.9$, we calculated the tail probability of seeing the observed number of successes or less.
If that probability was less than $10^{-3}$, then the simulation was regarded as rejecting the tested value of $M$, and confirming it otherwise.
Monte-Carlo simulation was carried out for all instances and all angles, and has never rejected the calculated value of $M$.
Thus, the results of Figure~\ref{fig:shots-vs-gens} were deemed reliable.

While this establishes the shot count needed to estimate the AR expectation to fixed accuracy for the system sizes considered here, related studies suggest that the precision required for successful variational optimization may become more demanding with increasing problem size under finite-shot or stochastic loss noise \cite{scriva2024challenges, barligea2025scalability}, motivating non-variational approaches that avoid repeated noisy loss evaluations \cite{catli2025efficient}.
In the results following below, the number of shots was chosen as $M = 1000$ for all tests, which is more than sufficient according to Figure~\ref{fig:shots-vs-gens}. 

\subsection{Solvers}
\label{sec:solvers}

\begin{figure}
    \centering
    \includegraphics[width=\linewidth]{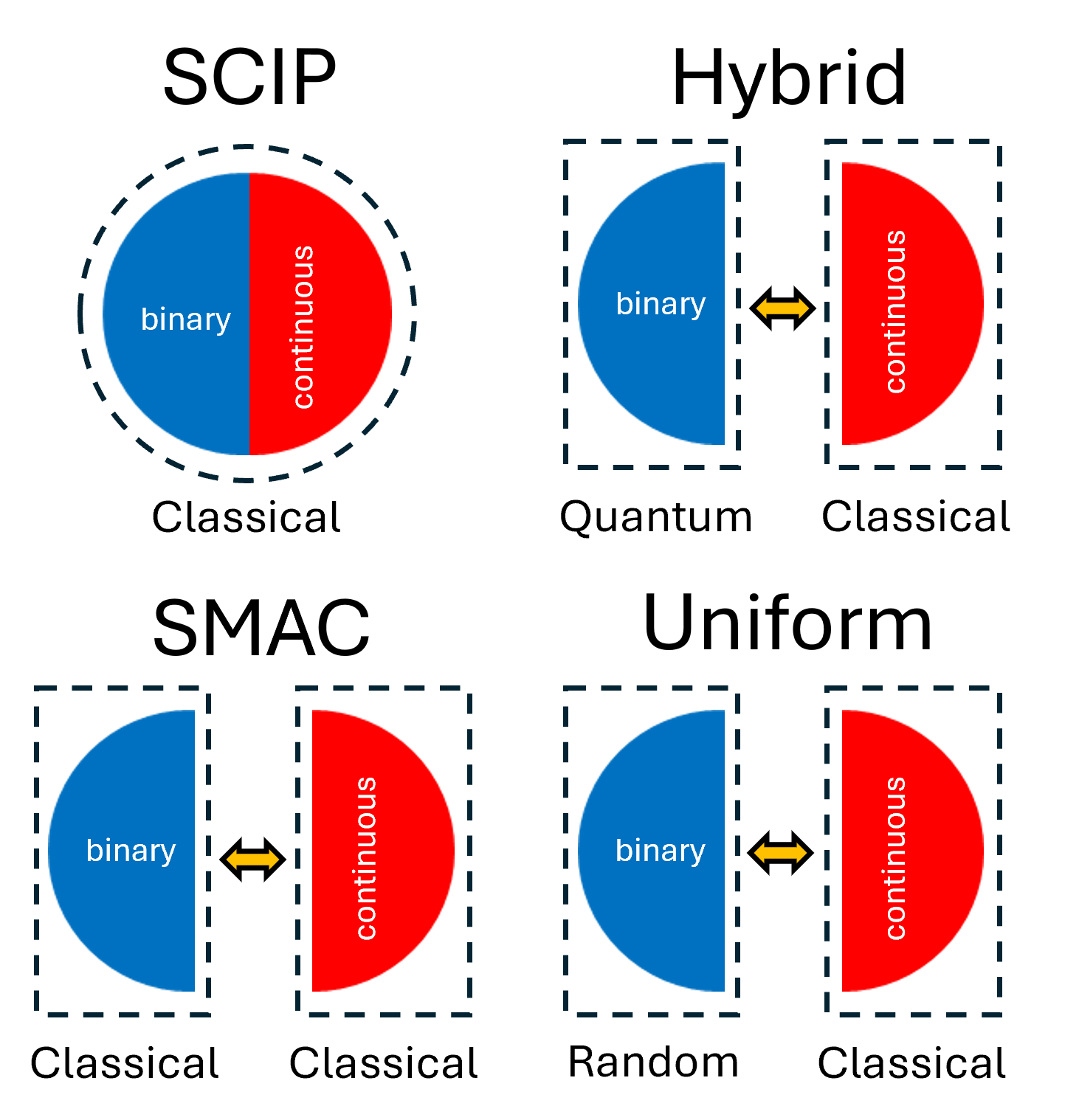}
    \caption{Schematic representation of the AC-OPF-UC solvers considered in this work. Binary and continuous variables for SCIP are optimized together with a single mixed integer optimizer. For other solvers they are optimized separately with 2 different optimizers indicated at the bottom.}
    \label{fig:solvers}
\end{figure}

In order to establish whether the proposed algorithm of Section~\ref{sec:qubit-efficient-algorithm} makes sense, we compared its performance with alternative ways to solve the problem.
An overview of the solvers considered in this benchmark is presented in Figure~\ref{fig:solvers}.

The most straightforward approach is to solve the problem by describing it exactly as stated to a classical mixed integer optimizer, without separating the variables into binary and continuous.
This baseline was implemented using a mixed integer solver library called \textbf{SCIP} \cite{bestuzheva2023scip}.

SCIP imposed all constraints explicitly and used a feasibility tolerance of $10^{-10}$, feasibility emphasis, aggressive heuristics, and fast presolving and separation. Exact implementation details are available in the public code repository.

Another approach is to act according to the prescription of the algorithm of Section~\ref{sec:qubit-efficient-algorithm}, i.e. by separating binary and continuous variables, optimizing continuous variables classically for fixed binary assignments, and feeding the result of this optimization to a separate optimization layer over the binary variables performed quantumly via VQA with a noisy classical optimizer.
This approach will be labeled as \textbf{Hybrid}.

Alternatively, we may decide not to involve quantum computer at all and optimize binary variables classically too.
In this benchmark, this approach was implemented using a Bayesian optimization library for binary variables called \textbf{SMAC} \cite{lindauer2022smac3}.
Other than that, everything else is the same as in Hybrid.

Another alternative is to not optimize binary variables at all, and just keep sampling them randomly with a uniform probability distribution until we run out of the optimization time budget.
This approach will be labeled as \textbf{Uniform}.
Once again, everything else is the same as in Hybrid and SMAC.

In the case of Hybrid, we tried multiple different noisy optimizers for VQA, including noisyopt \cite{mayer2017noisyopt}, SPSA \cite{spall2002multivariate}, Adam \cite{kingma2014adam}, and QNSPSA \cite{gacon2021simultaneous} with different options.
In our experience, ADAM with custom SPSA-style gradient demonstrated the best performance and was chosen as the noisy optimizer for the following tests. 
The following non-default settings have been used: learning rate = 0.04, SPSA perturbation = 0.1, AMSGrad enabled.

The value of $\lambda$ used in Eq.~\eqref{eq:cost} was chosen as 10 for all solvers.
In all solvers except SCIP, the continuous subproblem for each fixed bitstring was solved using CasADi \cite{andersson2018casadi} with IPOPT \cite{wachter2006implementation}, with the original constraints imposed explicitly and default settings. 
The penalized cost of Eq.~\eqref{eq:penalized-cost} was evaluated only after the inner optimization and passed to the outer optimizer. 
IPOPT was initialized with generator powers at the midpoints of their bounds, voltage magnitudes equal to 1, and voltage angles equal to 0. 
Results were cached by bitstring. 
Each new bitstring was given a 30-second time limit. 
On timeout or solver failure, the best candidate found was retained, with any constraint violation included in its penalized cost.

In all tests below, the time budget for any given system size was chosen as large enough to finish variational angle optimization for the Hybrid method on at least 100 out of 120 instances of a given dataset.
The slowest 20 instances were discarded and the same retained 100 instances with the same time budgets were given to other solvers.

The reason for this discard is that our broad parameter distributions were occasionally producing atypically hard instances taking much longer to converge.
Since their number and contribution to the overall set average were small, they were neglected.
While skipping the instances that are hard for the Hybrid method can in principle bias the subset selection in its favor, in practice instance hardness was largely shared across solvers.
Moreover, any potential bias in favor of Hybrid would not change our conclusions below.

\begin{figure}
    \centering
    \includegraphics[width=\linewidth]{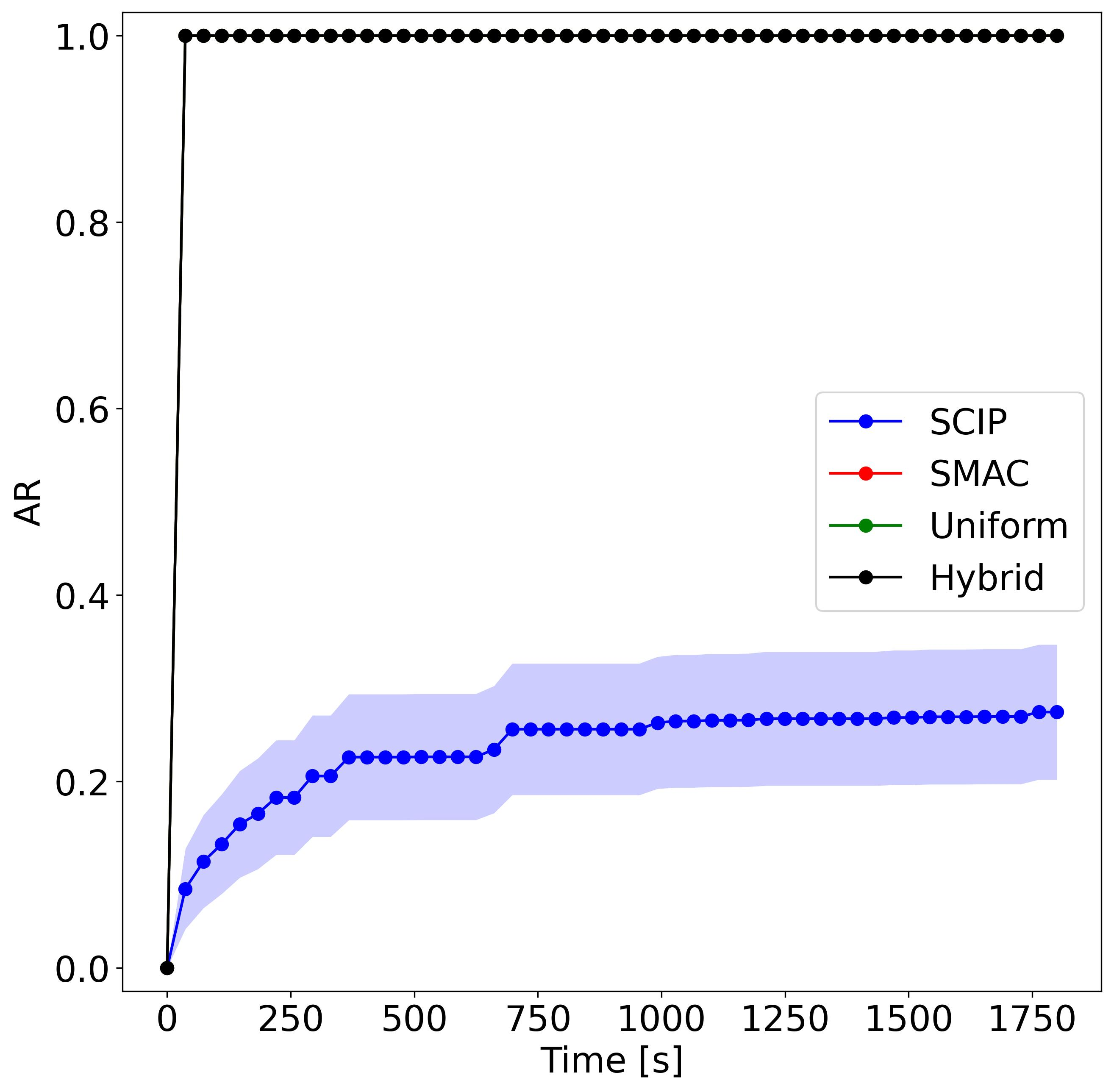}
    \caption{Instance-averaged approximation ratio of the best incumbent solution vs time for the 5 generators dataset. SMAC, Uniform and Hybrid all have exactly the same AR = 1. Shaded area shows 90\% confidence interval for individual data points.}
    \label{fig:5-gen}
\end{figure}

In the first benchmark, we compared all 4 approaches on a 5 generator dataset.
All methods were given 30 minutes per instance time budget.
During the optimization we recorded the time of finding each improvement over the previous best known solution found so far, thus obtaining AR vs time curve for each instance.
Then we averaged those curves over the retained 100 instances in the dataset.
The results of it are shown in Figure~\ref{fig:5-gen}.

As one can see, SMAC, Uniform and Hybrid all finished enumerating all bitstring assignments within seconds and found the best observed solution (AR = 1) for all instances, which is unsurprising considering the small system size of only 5 generators ($2^5=32$ possible assignments).
Their confidence interval widths are exactly 0.
On the other hand, SCIP was running for the entire 30 minutes time budget, but was still not able to find any feasible solutions for most of the instances in the dataset, which counts as AR = 0, thus earning its low instance-averaged AR.
For this reason SCIP was deemed non-competitive for this benchmark and was excluded from the other tests below.

One important caveat regarding the Hybrid's accounting in these benchmarks is that the quantum part (construction of quantum state for a given set of variational angles) was simulated.
It would not be fair to directly add quantum simulation time to the overall method time, because if we were to use an actual quantum hardware instead of simulation, the quantum state could have been built much faster.
But how much faster is hard to say because it depends on the specific hardware architecture \cite{krantz2019quantum, bruzewicz2019trapped, kuanysheva2026performance}.
In these benchmarks we decided to consider the ideal asymptotic case of infinitely faster, not counting the time spent on quantum simulation towards the total time at all.

\begin{figure}
    \centering
    \includegraphics[width=\linewidth]{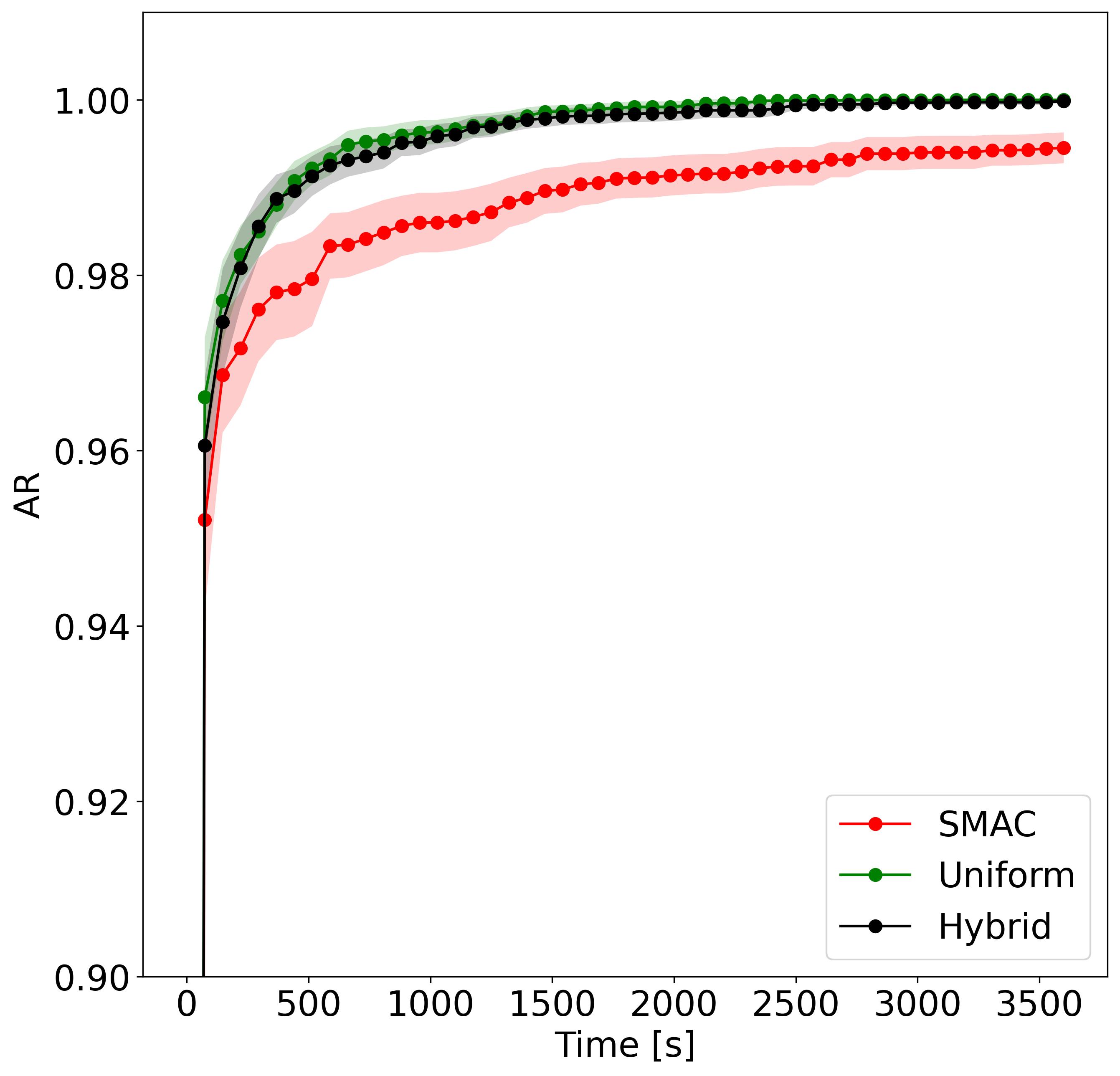}
    \caption{Instance-averaged approximation ratio of the best incumbent solution vs time for the 13 generators dataset. Shaded area shows 90\% confidence interval for individual data points.}
    \label{fig:13-gen}
\end{figure}

Figure~\ref{fig:13-gen} shows a zoomed-in instance-averaged AR vs time curves for the 13 generators dataset.
As one can see, all 3 remaining solvers still demonstrate nearly identical performance, but SMAC starts to fall a bit behind the other 2 solvers.
For this reason, in the following comparison we focus solely on the 2 leading solvers: Hybrid and Uniform.

\begin{figure}
    \centering
    \subfloat{\includegraphics[width=0.5\linewidth]{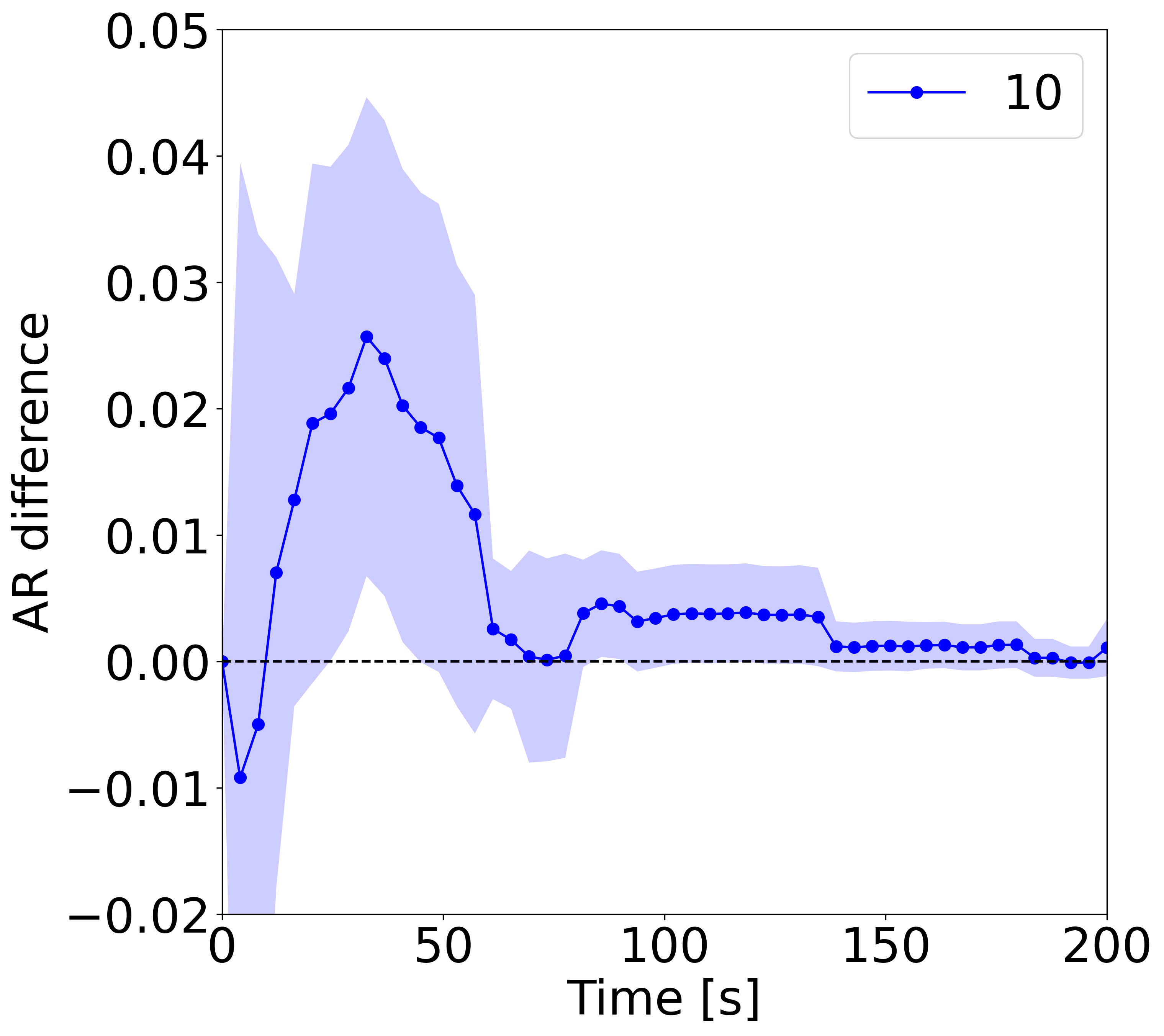}}
    \hfill
    \subfloat{\includegraphics[width=0.5\linewidth]{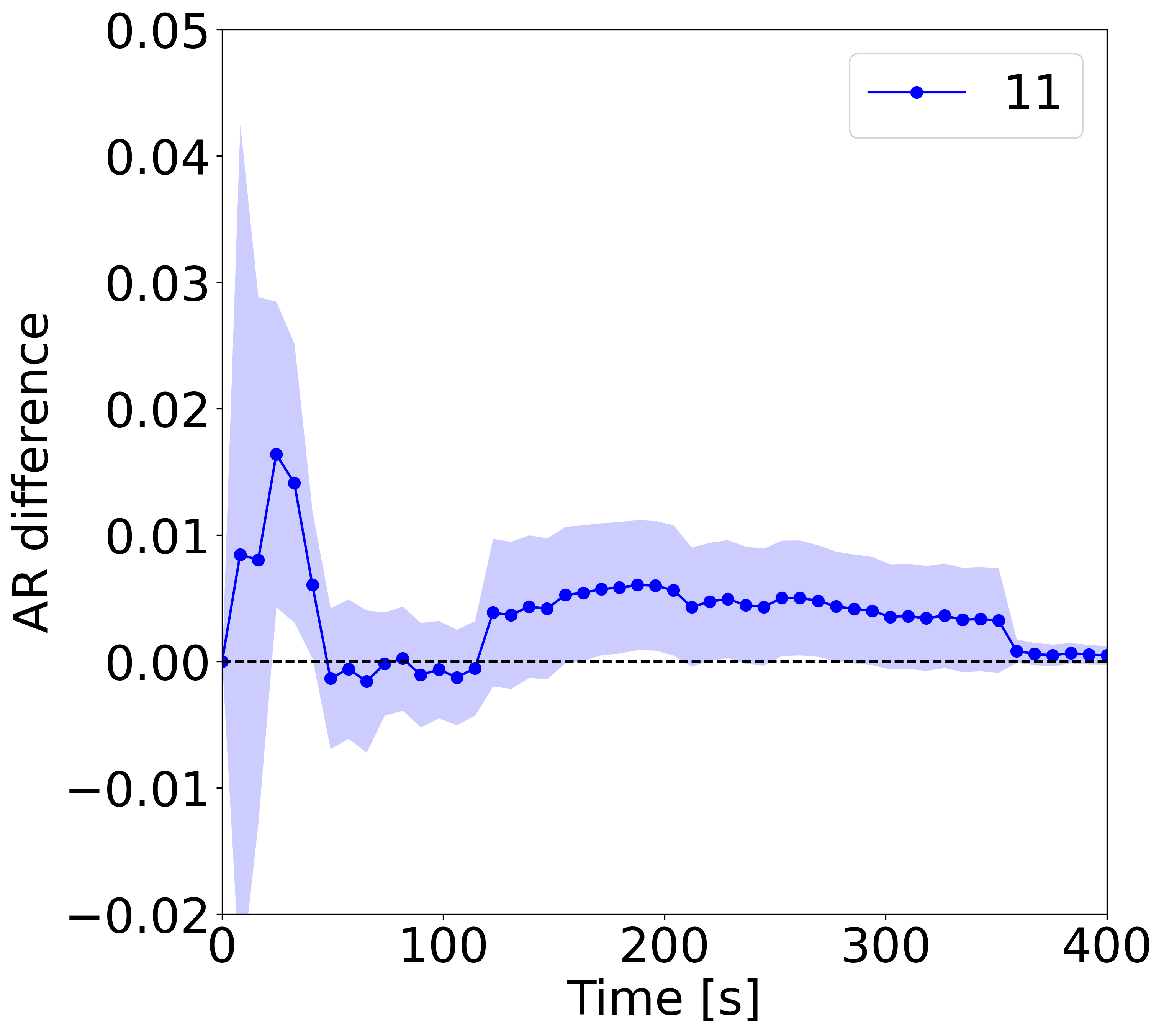}}

    \subfloat{\includegraphics[width=0.5\linewidth]{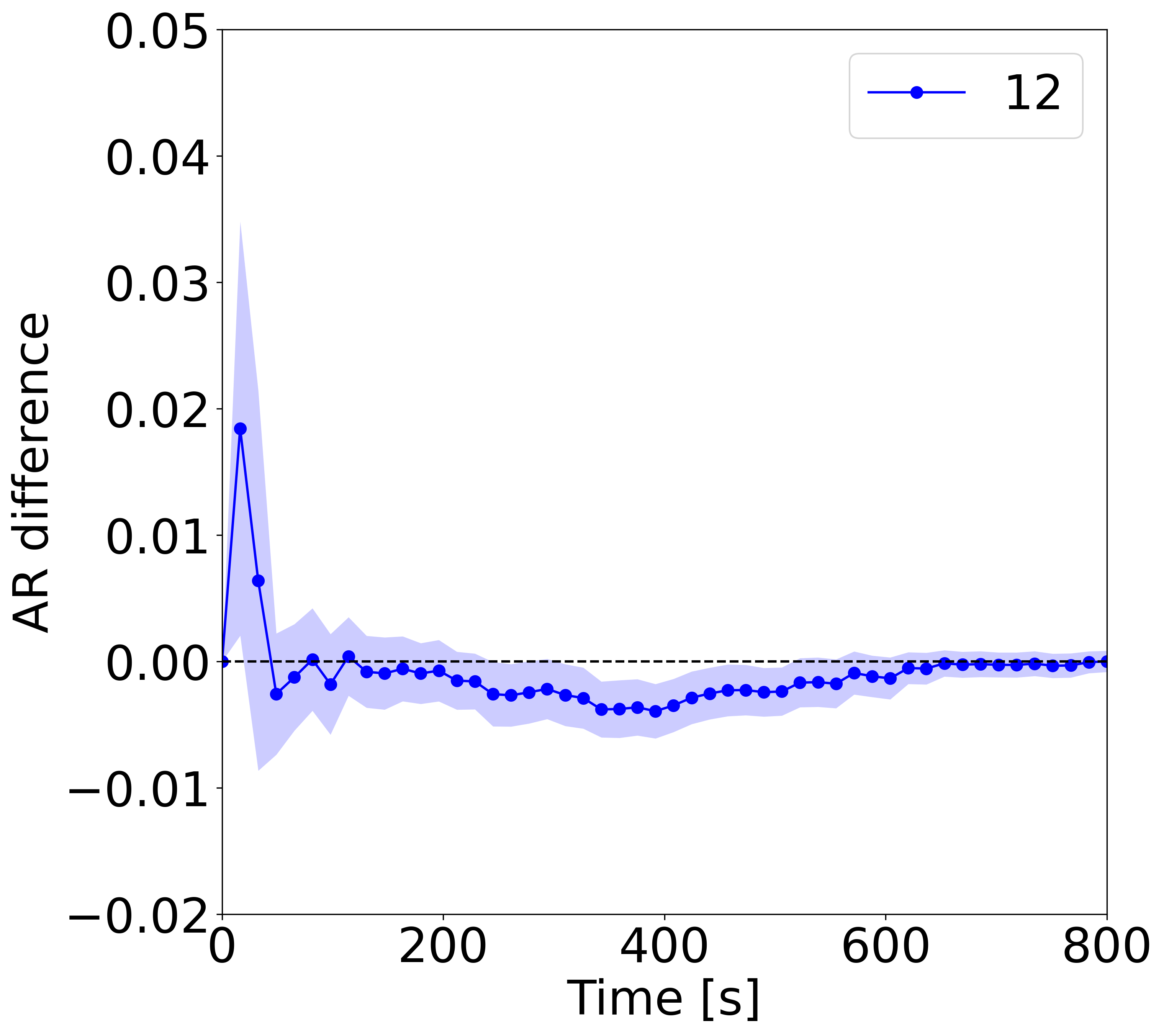}}
    \hfill
    \subfloat{\includegraphics[width=0.5\linewidth]{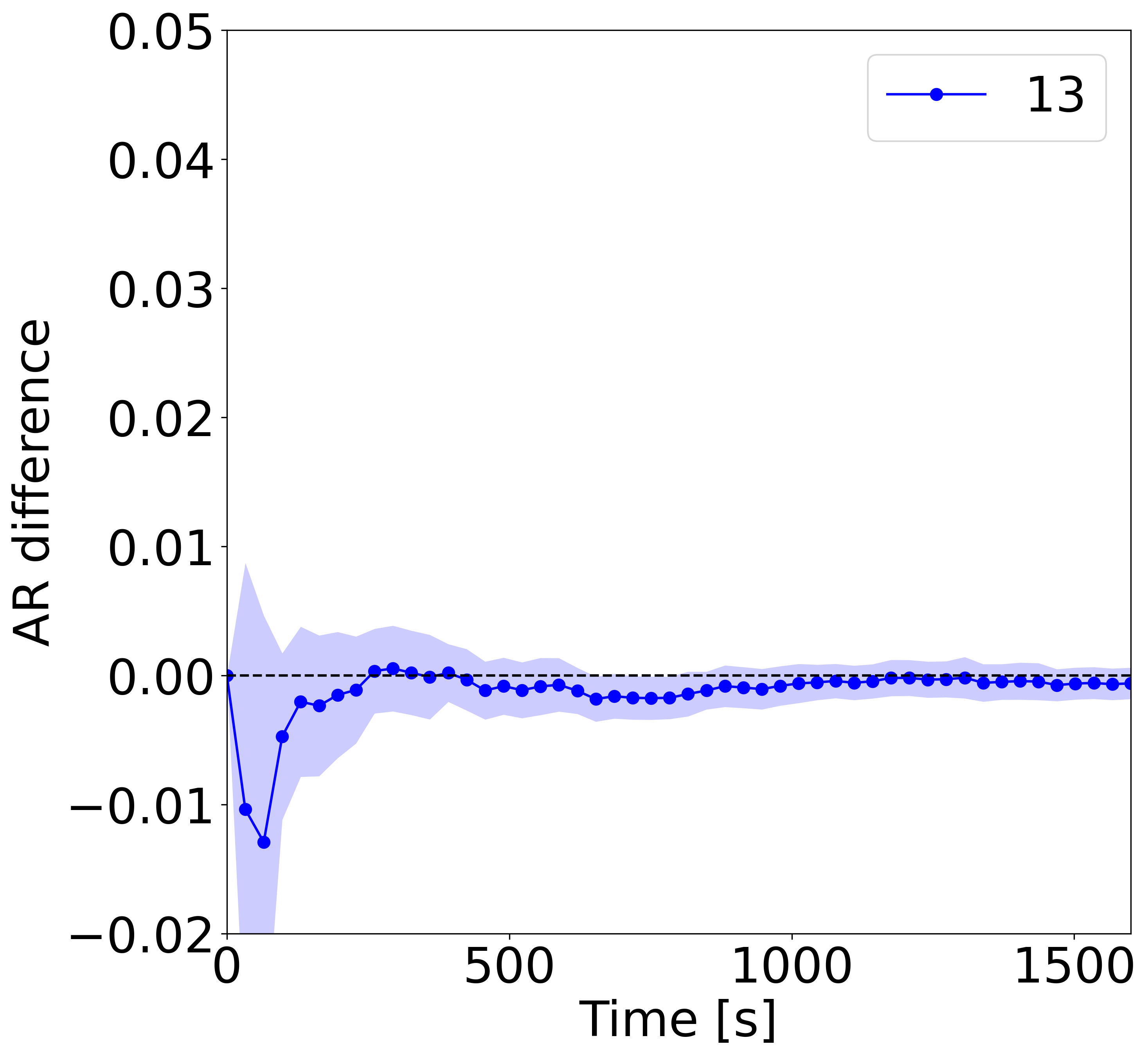}}
    \caption{AR of Hybrid minus AR of Uniform vs time for number of generators ranging from 10 to 13. Shaded area shows 90\% confidence interval for individual data points.}
    \label{fig:history-diff}
\end{figure}

In Figure~\ref{fig:history-diff}, we plot the difference between ARs (Hybrid minus Uniform) vs time for number of generators ranging from 10 to 13.
Asymptotically, in the limit of infinite time, there will be no difference between the methods since all bitstrings will be sampled one way or another and the same best solution will be found.
In the limit of zero time, we also do not expect to see meaningful difference because VQA optimization starts from random angles, which are not expected to create a probability distribution better than uniform, on average.
As one can see from Figure~\ref{fig:history-diff}, intermediate times also do not demonstrate large enough differences to matter in practice.

The main reason for this effect is that the considered system sizes are too small.
With $M = 1000$ shots per sample mean, all or most of their bitstring configurations will be sampled during the first few optimization iterations, while the variational angles are still close to the initial random angles.
By the time variational angle optimization achieves meaningful separation between the optimized and uniform probability distributions, the best bitstring will have long been found, and any further optimization will not have any effect on the AR vs time curves.

For example, 13 generator instances require $\approx\!\!10000$ iterations to finish outer optimization.
With $M = 1000$ shots per iteration, they sample roughly $\approx10^7$ bitstrings during optimization.
The smallest system that even has this many bitstrings is 24 qubits.

As system size increases, the total number of possible bitstrings increases as well (exponentially), so it will take longer to sample the high-AR bitstring randomly, which should give optimization more time to become meaningful.
However, this effect is partially offset by the fact that the number of iterations necessary to finish optimization also increases with system size, albeit not as quickly as number of bitstrings (from $\approx$5000 iterations at 10 generators to $\approx$10000 iterations at 13 generators on average in our dataset, increased approximately linearly).

\begin{figure}
    \centering
    \subfloat[10 generators]{\includegraphics[width=\linewidth]{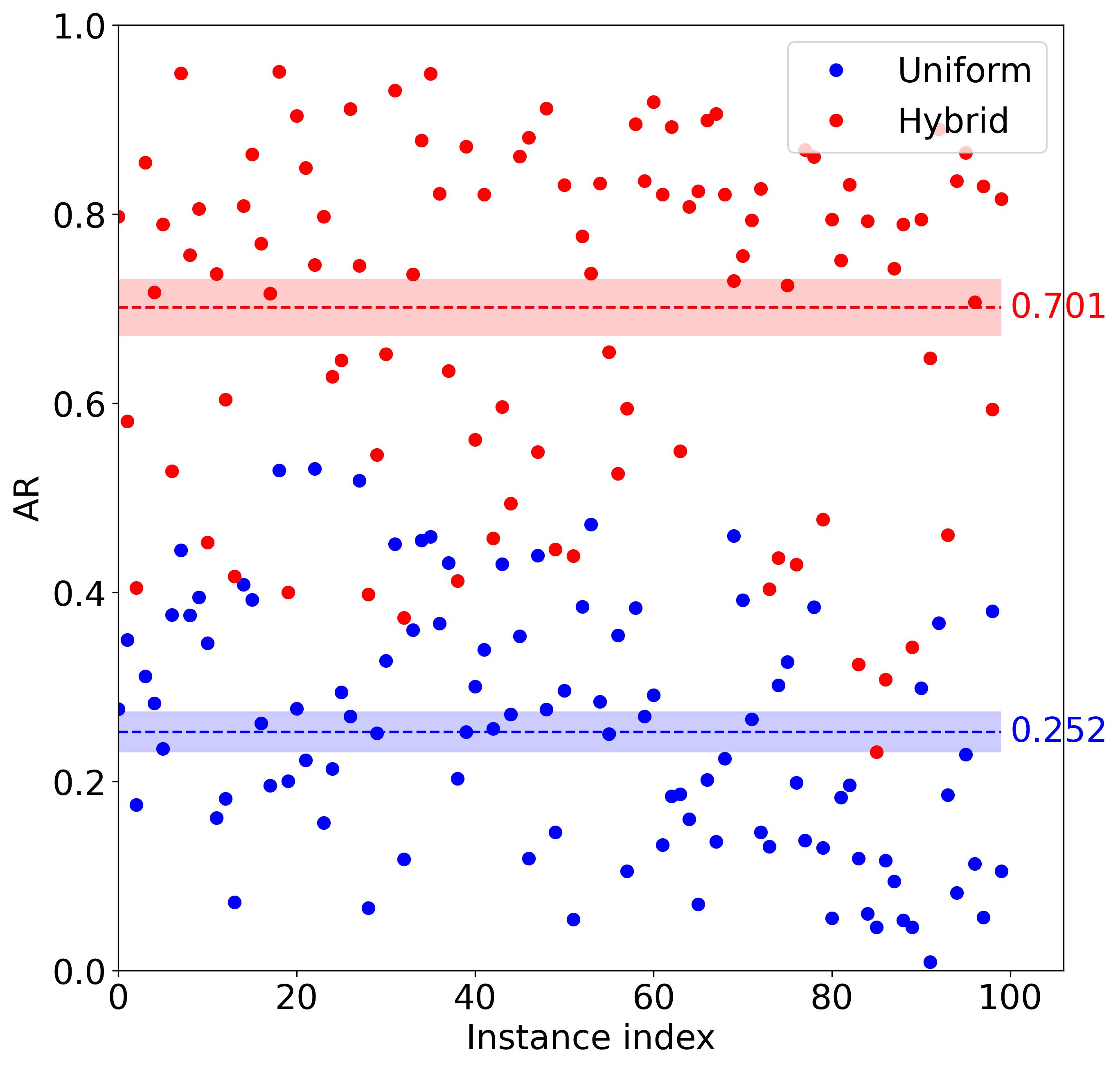}}

    \subfloat[13 generators]{\includegraphics[width=\linewidth]{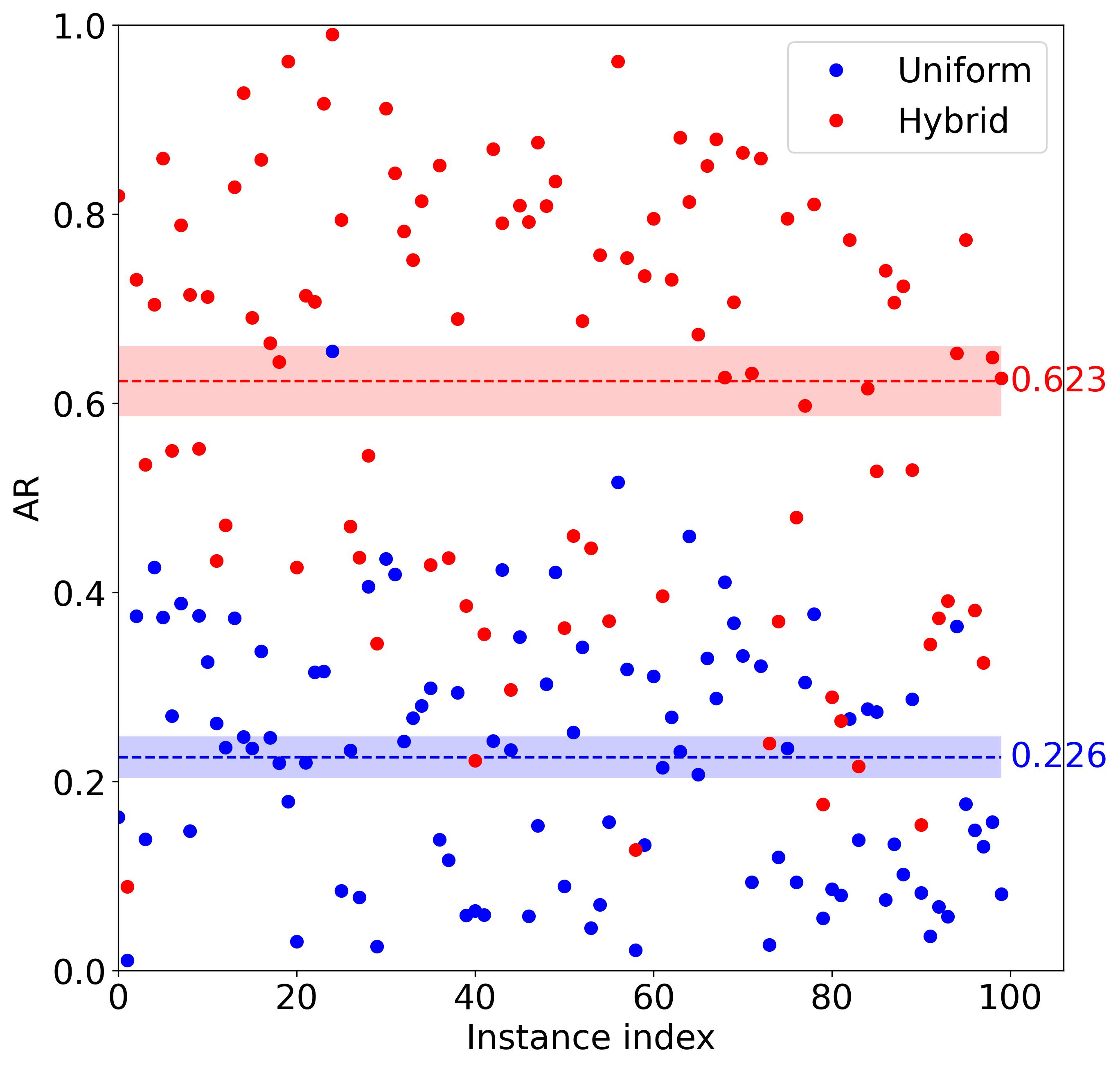}}
    \caption{ARs of individual instances for 10 and 13 generators case corresponding to Uniform and variational-angle-optimized (Hybrid) probability distributions over bitstrings. Instance index is just a sequential number assigned to each generated instance. Horizontal dashed lines show the instance-average values. Shaded area shows 90\% confidence interval for the means.}
    \label{fig:ar-vs-instance}
\end{figure}

Another issue is that performance of only one layer of the ansatz will continue to drop with system size, and it will do so faster than performance of uniform probability distribution, thus further bringing them closer together.
This effect is shown in Figure~\ref{fig:ar-vs-instance}.
Specifically, the figure shows expected values of AR corresponding to uniform and variational angle-optimized probability distributions for each individual instance in the 10- and 13-generator datasets.
The dashed lines show their average values.
As one can see, going from 10 to 13 generators, averaged optimized expectation drops by 0.078 ($\sim\!\!11\%$), while uniform expectation drops by 0.026 ($\sim\!\!10\%$).
While the relative drop is similar, the absolute value of the drop is a factor of 3 larger in the optimized case.

Adding more layers should help with that in principle, but it will further increase the number of iterations that optimization needs, and may also increase the number of shots necessary per iteration, which will further push the boundary of the smallest meaningful system size.
Moreover, AR gain with additional layers tends to follow diminishing returns \cite{gaidai2024performance, campos2021training}, requiring more and more layers to keep up with the system size.

Taking these issues into account, our previous estimation of 24 qubits will likely need to be increased even more.
But even for 24 qubits, assuming 10 seconds of inner optimization time per bitstring (extrapolated from our data), optimizing 100 instances would require around $5 \times 10^6$ CPU hours per solver, so over $10^7$ hours total, which is already beyond our computational capacity. 

Better than random initial angles are also not expected to help because our ansatz is independent of the cost function, so the same variational angles always translate to the same probability distributions regardless of instance, and the same probability distribution cannot possibly work well for all instances.

A tail-weighted objective, such as CVaR, could in principle improve convergence to the best configuration at large enough system size, but is unlikely to affect the conclusions of this study, since the search space gets exhausted before any substantial variational optimization can occur.

Note that while the quantitative results considered here are specific to the tested dataset, outlined in Section~\ref{sec:datasets}, we expect qualitatively similar behavior to hold for other benchmarks as well, including standard test systems, since the observed mechanisms, described above, depend primarily on the relative rates of discrete-space exploration and variational optimization rather than on particular instance distributions.

\section{Conclusions}
\label{sec:conclusion}

In this work we presented 2 candidate hybrid quantum-classical algorithms that can solve single-period AC-OPF-UC problem and investigated the performance of one of them with respect to the classical baselines.

The first method involved direct application of QAOA to evaluate the full cost function on a quantum computer, but that approach requires hundreds of qubits even for small system sizes.
For this reason we did not benchmark its performance.

The second approach attempts to fix the main issue of the first by separating the problem variables into binary and continuous, and optimizing only binary variables on a quantum computer, while continuous variables are optimized classically.
This dramatically decreases the number of required qubits, as intended.

However, numerical simulations show that this approach is not better than classical uniform random sampling, even in the maximally quantum-favorable case of assuming that all quantum operations take 0 time.
At least not for the considered system sizes.
For the reasons discussed at the end of Section~\ref{sec:solvers}, much larger system sizes (24+ generators, likely more) need to be tested before the hybrid algorithm even has a chance to outperform uniform sampling.

We estimate that such system size would need roughly $\approx\!10^7$ CPU hours to be performed consistently with our current benchmarking methodology, which is beyond our computational capacity.
The same benchmarking procedure can be straightforwardly adapted to DC and UC problems as well.

Future directions may include the following:
\begin{enumerate}
    \item Even without evaluating the true cost function on quantum computer, an ansatz can be made instance-dependent by incorporating some reasonable proxy (surrogate) cost function into it.
    This could potentially enable angle transfer approaches \cite{shaydulin2023parameter, galda2023similarity, jing2023data} or meaningful angle initialization strategies \cite{zhou2020quantum, sack2021quantum, lee2023depth, gaidai2024performance, kim2026safe}.
    
    \item Different ansatzes or training schemes could potentially reduce the required number of optimization iterations \cite{herrman2022multi, campos2021training, lee2024iterative, jang2026cyclic}.
    
    \item Other methods, such as quantum branch and bound methods, Benders decomposition, Lagrangian relaxation, ADMM or qudit-based approaches \cite{montanaro2020quantum, paredes2021benders, tuncer2022misocp, feng2022novel, magar2024dc, goswami2024integer, goswami2025qudit} are likely more promising.

    \item A classical trainable probabilistic sampler could help determine whether the observed behavior is specific to the quantum variational model or is a more general limitation of adaptive probability-distribution learning.
\end{enumerate}

\section*{List of abbreviations}

\begin{description}
    \item[AC] Alternating current
    \item[ADMM] Alternating direction method of multipliers
    \item[AR] Approximation ratio
    \item[CDF] Cumulative distribution function
    \item[DC] Direct current
    \item[HHL] Harrow--Hassidim--Lloyd algorithm
    \item[IPOPT] Interior Point Optimizer
    \item[KKT] Karush--Kuhn--Tucker
    \item[MILP] Mixed-integer linear programming
    \item[MINLP] Mixed-integer nonlinear programming
    \item[MIQCQP] Mixed-integer quadratically constrained quadratic programming
    \item[MIQP] Mixed-integer quadratic programming
    \item[MISOCP] Mixed-integer second-order cone programming
    \item[OPF] Optimal power flow
    \item[QAOA] Quantum approximate optimization algorithm
    \item[QNSPSA] Quantum natural simultaneous perturbation stochastic approximation
    \item[QUBO] Quadratic unconstrained binary optimization
    \item[RMS] Root mean square
    \item[SCIP] Solving Constraint Integer Programs
    \item[SDP] Semidefinite programming
    \item[SMAC] Sequential Model-based Algorithm Configuration
    \item[SOCP] Second-order cone programming
    \item[SPSA] Simultaneous perturbation stochastic approximation
    \item[UC] Unit commitment
    \item[VQA] Variational quantum algorithm
\end{description}

\section*{Declarations}

\subsection*{Availability of data and materials}
For the exact code, finer technical details, and the raw data supporting the findings of this study, the readers are invited to inspect our GitHub repository:
\url{https://github.com/GaidaiIgor/PowerGridQuantumOpt}.

\subsection*{Competing interests}
The authors declare that they have no competing interests.

\subsection*{Funding}
This work was supported by Tennessee Valley Authority (TVA) and by NIST through the CIPP program under award 60NANB24D218.
The funders had no role in the design of the study, analysis, interpretation of results, or writing of the manuscript.

\subsection*{Authors' contributions}
IG developed the algorithms, performed the numerical simulations, analyzed the results, and wrote the manuscript.
RM acquired funding, supervised the project, contributed to the interpretation of the results, and revised the manuscript.
Both authors read and approved the final manuscript.

\subsection*{Acknowledgements}
The authors thank James Cummins for his help with the literature survey.

\subsection*{Use of generative artificial intelligence}
The authors used OpenAI GPT-based generative artificial intelligence tools to assist with code development and debugging, study design and methodology, literature-review organization, manuscript drafting and editing.
All scientific decisions, code, analyses, interpretations, conclusions, and final manuscript text were reviewed, verified, and approved by the authors.

\bibliographystyle{unsrt}
\bibliography{refs}

\end{document}